\documentclass[english,aps,superscriptaddress,amsmath,amssymb,longbibliography,onecolumn,floatfix,10pt]{revtex4-1}
\usepackage{amsmath}
\usepackage{graphicx}  
\usepackage{dcolumn}   
\usepackage{bm}        
\usepackage{amssymb}   
\usepackage[normalem]{ulem}

\usepackage{hyperref}
\hypersetup{
    colorlinks=true,
    linkcolor=blue,
    filecolor=blue,      
    urlcolor=black,
    citecolor=blue,
        }
\usepackage{subfigure}
\begin{document}

\title{  Motility and Swimming: Universal Description and Generic Trajectories }
\author{Alexandr Farutin}
\affiliation{Univ. Grenoble Alpes, CNRS, LIPhy, F-38000 Grenoble, France}
\author{Suhail M. Rizvi}
\affiliation{Univ. Grenoble Alpes, CNRS, LIPhy, F-38000 Grenoble, France}
\affiliation{Current address: Department of Biomedical Engineering, Indian Institute of Technology Hyderabad, Sangareddy, Telangana 502285, India}
\author{Wei-Fan Hu}
\affiliation{Department of Mathematics, National Central University, 300 Zhongda Road, Taoyuan 320, Taiwan}
\author{Te-Sheng Lin}
\affiliation{Department of Applied Mathematics, National Chiao Tung University,
1001 Ta Hsueh Road, Hsinchu 300, Taiwan}
\author{Salima Rafai}
\affiliation{Univ. Grenoble Alpes, CNRS, LIPhy, F-38000 Grenoble, France}
\author{Chaouqi Misbah}
\email[]{chaouqi.misbah@univ-grenoble-alpes.fr}
\affiliation{Univ. Grenoble Alpes, CNRS, LIPhy, F-38000 Grenoble, France}


\date{\today}

\begin{abstract}
Autonomous locomotion is a ubiquitous phenomenon in biology and in  physics of active systems at microscopic scale. This includes prokaryotic, eukaryotic cells  (crawling and swimming) and artificial swimmers. An outstanding feature is the ability of these entities to follow complex trajectories, ranging from straight, curved (circular, helical...), to random-like ones.
The non-straight nature of these trajectories is often explained as a consequence of the asymmetry of the particle or the medium in which it moves, or due to the presence of bounding walls, etc...
Here, we show that straight, circular and helical trajectories emerge naturally in the absence of asymmetry of the swimmer or that  of suspending  medium. Our first proof is based on general considerations, without referring to  an explicit form of a model. We  show that these three trajectories  correspond to self-congruent solutions.
Self-congruency means that the states of the system at different moments of time can be made identical by an appropriate combination of rotation and translation of the coordinate space.
We show that these solutions are exhibited by spherically symmetric particles as a result  of a series of pitchfork bifurcations as the activity is increased.
Self-congruent dynamics in one and two dimensions are analyzed as well.
Finally, we present a simple explicit nonlinear  exactly solvable model of fully isotropic phoretic particle that shows the transitions from a non-motile state to straight motion to circular motion to helical motion as a series of spontaneous symmetry-breaking bifurcations.  Whether a system exhibits or not a given trajectory only depends on the numerical values of parameters entering the model, while asymmetry of swimmer shape, or anisotropy of the suspending medium , or influence of bounding walls are not necessary.
\end{abstract}
\maketitle

\section{Introduction}
Microorganisms, such as  eukaryotic and prokaryotic cells, as well as as their biomimetic counterparts, such as active particles and drops, constitute today a fertile interdisciplinary topic.
Consumption of energy and production of entropy enabling the motility make the traditional concepts of equilibrium statistical mechanics  difficult to apply.
These systems show a wide range of unexpected behaviors regarding  both their individual and collective behaviors, such as complex trajectories, non trivial patterns, and so on. 

It is recognized since more than a century that living  entities (flagellate, spores, infusoria, and so on) have the ability to follow complex trajectories, such as spiral ones\cite{Jennings1901}.
Other trajectories, like circular, helical, and even chaotic have been later identified for living cells and artificial particles\cite{Shenoy2007,Riedel2005,jana2012paramecium,kruger2016curling,Loewen2016,suga2018self, narinder2018memory,izri2014self,Hu2019}.
These systems may combine intricate biochemical regulations, chemical and mechanical interactions,  and nonlinear effects making their explicit elucidation quite challenging.

Cells can swim in a fluid, crawl on a substratum or move in a complex environment (extracellular matrix, complex fluid, random media...) by using several strategies, such as shape change (amoeboid motion), beating of cilia and flagellae, and so on\cite{marchetti2013hydrodynamics,Lauga2009,LaugaBJ,barry2010dictyostelium,ONEILL2018,Theodoly}.
The motility can also be powered by a chemical field.
Prominent examples of the latter are Marangoni-driven particles\cite{MLB13,morozov2019nonlinear,schmitt2013swimming}, and acto-myosin assisted cell motility\cite{recho2013contraction,Recho2015a,hawkins2011spontaneous,Voituriez2016,Farutin2019}.
Another related phenomenon is the formation and motion of concentration peaks in bulk active medium\cite{Negro2019}, which was also observed in epithelial tissue during embryo growth\cite{BLANCHARD201878}. Finally, similar problem arises in the study of active phase field crystals\cite{Menzel2013,Menzel2014,Ophaus2018}, where the patterns of the phase field can be used to model swarms of active particles.

The interplay between various ingredients  involved in cell  motility and swimming poses often  a formidable challenge in terms of effective basic description and classification of their dynamics.
At the same time it is expected  that these systems should obey some basic rules  when formulated  in  a general framework based on symmetries and  non equilibrium  constraints (such as the lack of a variational formulation), regardless of the specific nature of the motile system under consideration.   
 
Our focus here is the understanding of the emergence of generic curved trajectories.
The occurrence of complex curved trajectories has been linked to the geometry of the particle (such as its chirality)\cite{Loewen2016}, or to the nature of the suspending medium (complex fluid)\cite{narinder2018memory}, or to noise \cite{MAIURI2015374}, or to the presence of bounding walls \cite{LaugaBJ}.
This has led to interesting descriptions of their motion.
Surprisingly, complex motions have also been reported for isotropic particles (circular, spherical) where circular, meandering and chaotic  motions are revealed in the absence of noise and bounding walls\cite{Hu2019}, pointing to their genericity.  

The present study focuses on the description of generic motions of a single entity driven by a chemical field (extensions to several fields will appear straightforward), be it crawling on a substrate, or moving in a fluid, without a specific reference to a system or to the environment in which it moves.
By introducing  a simple but a general framework where a particle is driven by a concentration field we will be able to state the general conditions under which a particle can transition from a non-motile to a motile state, and will identify the nature of the resulting bifurcation.
We will identify the existence of generic types of motion that we will refer to as {\it self-congruent solutions}.
These are solutions that keep the concentration field moving and rotating in a shape-preserving manner in the course of time.
They correspond to three types of solutions: (i) straight trajectory, (ii) circular one, and (iii) helical ones.
These solutions are identified on the basis of general symmetry properties.  
 
The next step consists in writing explicitly, again based on symmetry, the evolution equations of the concentration fields components.
It will appear that self-congruent solutions emerge naturally from explicit solutions of nonlinear equations.
We will keep the equations simple enough, and we will show that retaining only two components (first two  spherical harmonics) is sufficient to yield a complex picture.
It turns out that these equations, though nonlinear, yield an exact solution describing the above three trajectory types.
We will discuss here three  motile systems, namely a segment in 1D, a circular particle in 2D and  a spherical particle in 3D.
This study highlights the potential of this framework which captures the main three essential motions in a unified way.
In the companion Letter \cite{Farutin2020PRL} we also show that these equations yield chaotic motions.
We will explain how this study can help to analyze efficiently explicit models of motility.


















\section{Problem formulation}
The purpose of this work is to analyze the trajectories of self-propelling particles and the bifurcations which lead to transition from one trajectory type to another.
We aim to consider this problem as generally as possible making little assumptions about the nature of the particle or the properties of the medium in which it moves.
For this purpose, we assume that the motility of the particle is related to a concentration field $c$ distributed on the particle surface and, possibly, in the bulk media inside or outside the particle. Typical examples include Marangoni-driven particles \cite{MLB13,morozov2019nonlinear,schmitt2013swimming}, and acto-myosin assisted  cell motility \cite{hawkins2011spontaneous,Voituriez2016,Farutin2019}.

The equations governing the phoretic dynamics can be formally written as (see below for an explicit example)
\begin{equation}
\label{functional}
\dot c(\boldsymbol r,t)=\mathcal G\{c(\boldsymbol r',t)\}(\boldsymbol r),
\end{equation}
where $\mathcal G\{c(\boldsymbol r')\}(\boldsymbol r)$ is a non-linear operator and the dot refers to partial derivative with respect to time.
The operator $\mathcal G$ takes some function $c$ of coordinate $\boldsymbol r'$ and returns another function of $\boldsymbol r'$, which we denote as $\mathcal G\{c(\boldsymbol r')\}$.
Sampling this function at a given point $\boldsymbol r$ is denoted as $\mathcal G\{c(\boldsymbol r')\}(\boldsymbol r)$.
The value of $\dot c$ at some point $\boldsymbol r$ and time $t$ is a non-linear function of the values of $c$ at all possible points $\boldsymbol r'$ and the same time $t$. 
In other words, the evolution equation (\ref{functional}) is nonlocal in space but local in time.

Since the problem we consider here is local in time, the dependence on $t$ is implied below, whenever such simplification of notations does not lead to confusion.
The notation (\ref{functional}) codifies that the evolution of concentration can be unambiguously reconstructed from the instantaneous concentration field.
This is thus the most general form of the governing equation for an autonomous self-propulsion problem.

Let us consider an explicit model studied in \cite{MLB13,morozov2019nonlinear,schmitt2013swimming,Hu2019}.
This is an autophoretic particle which can emit (or absorb) a chemical substance at its surface. 
In that case the tangential velocity field along the particle is related to the surface gradient of the chemical concentration.
The flow equations can be solved for to express the fluid velocity in terms of the concentration field at the particle surface.
The concentration dynamics in the bulk, which obeys an advection-diffusion equation, defines a nonlinear evolution equation for the concentration in the bulk and at the surface \cite{MLB13,morozov2019nonlinear,schmitt2013swimming,Hu2019}.
For example, the evolution equation reads
\begin{equation}
\label{Eq:advection-diffusion}
\begin{aligned}
\dot c(\boldsymbol r,t)&= \sum _k { c_k(1,t) M e^{ik\phi } \over 2 r^{\vert k\vert +1} }  \left[ (r^2-1)k^2 \frac{\partial c (\boldsymbol r,t) }{\partial r}  +ik(2r^2 +(1-r^2) \vert k\vert ) \frac{\partial c (\boldsymbol r,t)}{\partial \phi} \right]+ \frac{1}{Pe}\Delta c (\boldsymbol r,t) 
\\&\equiv \mathcal G\{c(\boldsymbol r',t)\}(\boldsymbol r)
\end{aligned}
\end{equation}
in 2D for a circular particle.
This is a nonlocal equation since the evolution of the field $c$ at the considered point $\boldsymbol r$ in the bulk depends on concentration elsewhere:
the knowledge of the field along the bead surface  at $r=1$ and at infinitesimally close points.
The latter information is needed to calculate the derivatives of the concentration field.
Here $c_k$ refers to $k$-th Fourier component, $\phi$ is the polar angle, and  $M$ is a real constant relating the tangential velocity of the fluid and the gradient of $c$ along the bead \cite{MLB13,morozov2019nonlinear,schmitt2013swimming,Hu2019}.
The particle velocity is calculated as $(-M\Re c_1(1,t),M\Im c_1(1,t))$.
This defines the full evolution of the concentration in the laboratory frame.

The nonlinear operator $\mathcal G$ in eq. (\ref{functional}) must respect the symmetry of the underlying space.
That is, for any orthogonal matrix $\mathsf R$ and translation vector $\boldsymbol r_0$,
\begin{equation}
\label{Gsym}
\mathcal G\{c(\mathsf R \boldsymbol \cdot \boldsymbol r'+\boldsymbol r_0)\}(\boldsymbol r)=\mathcal G\{c(\boldsymbol r')\}(\mathsf R\boldsymbol \cdot \boldsymbol r+\boldsymbol r_0).
\end{equation}

Since this work considers spontaneous symmetry breaking of the concentration field, the shape of the particle is taken to be invariant under rotations.
Namely, we consider the following 3 cases:
\begin{itemize}
\item 1D case: a segment particle on a line
\item 2D case: a circular particle in a plane
\item 3D case: a spherical particle in a space
\end{itemize}
In all cases the particle boundary can be considered {\bf fictitious}, taking a varying concentration field in an otherwise homogeneous space.
The particle position can be defined, for example, by the concentration maximum in this case.
Further below, the radius of the particle is taken as 1.
The particle speed ${\bf v}(t)$ can be expressed explicitly as a function of the concentration field \cite{MLB13,morozov2019nonlinear,schmitt2013swimming,Hu2019}.
The proposed formulation tracks the particle motion implicitly, by following the evolution of the concentration field.
The two approaches are equivalent for self-congruent solutions considered below.

We consider the symmetry breaking bifurcations which separate different motile states from each other and from the static case.
The bifurcations are expected to appear upon increase of the particle activity, which we denote as non-dimensional Peclet number $Pe$.

\section{\label{sect:SelfCongruent}Self-congruent solutions}
Self-similar solutions play an important role in many physical problems.
Along this line, we analyze a related class of solutions, which we call self-congruent solutions below.
We call a solution self-congruent if the concentration fields at different moments of time can be made identical by an appropriate composition of rotation and translation of the space (congruence transformation).
We choose the name by analogy with a widely used term of self-similar solution, for which the states at different moments of time can be made identical by a scaling transformation of the space.
We present in this Section the possible types of self-congruent solutions and analyze the resulting trajectories of the particle.
The next Section is dedicated to the discussion of how self-congruent solutions emerge from the non-motile state through a series of pitchfork bifurcations as the activity is increased.

The formal expression of self-congruency reads
\begin{equation}
\label{self-congruency}
c(\boldsymbol r,t)=C[\mathsf R(t)\boldsymbol \cdot (\boldsymbol r-\boldsymbol r_c(t))],
\end{equation}
where $\mathsf R(t)$ is a time-dependent rotation matrix and ${\bf r}_c(t)$ is a time-dependent { offset} vector. This means that  the concentration field moves in a shape-preserving manner, very much like the usual travelling-wave solution equation. This also means that two concentration fields at two different times can be made identical to each other via an appropriate translation and rotation.

We now substitute the ansatz (\ref{self-congruency}) into the dynamics equation (\ref{functional}):
\begin{equation}
\label{ansatz}
\begin{aligned}
[\dot{\mathsf R}(t)\boldsymbol\cdot(\boldsymbol r-\boldsymbol r_c(t))-\mathsf R(t)\boldsymbol\cdot\dot{\boldsymbol r}_c(t)]\boldsymbol\cdot\boldsymbol C'[\mathsf R(t)\boldsymbol \cdot (\boldsymbol r-\boldsymbol r_c(t))]&=\mathcal G\{C[\mathsf R(t)\boldsymbol \cdot (\boldsymbol r'-\boldsymbol r_c(t))]\}(\boldsymbol r, t)\\
&=\mathcal G\{C(\boldsymbol r')\}[\mathsf R(t)\boldsymbol \cdot (\boldsymbol r-\boldsymbol r_c(t))],
\end{aligned}
\end{equation}
where $\boldsymbol C'$ is the gradient of $C(\tilde{\boldsymbol{r}})$ with respect to its only vector variable $\tilde{\boldsymbol{r}}=\mathsf R(t)\boldsymbol \cdot (\boldsymbol r-\boldsymbol r_c(t))$.
The last equality in (\ref{ansatz}) is obtained by using the commutativity of the nonlinear operator $\mathcal G$ with rotations and translations of the space (\ref{Gsym}).
We now define a matrix $\mathsf\Omega(t)$, which is nothing but the angular velocity matrix of the corotating frame, expressed in the corotating frame,  such that 
\begin{equation}
\label{Omega}
\dot{\mathsf R}(t)=\mathsf\Omega (t)\boldsymbol\cdot\mathsf R(t)
\end{equation}
and a vector $\boldsymbol v(t)$ (which is the velocity of the particle in the corotating frame) such that 
\begin{equation}
\label{v}
\mathsf R(t)\boldsymbol\cdot\dot{\boldsymbol{r}}_c(t)=\boldsymbol v(t).
\end{equation}
Substituting these definitions into (\ref{ansatz}) yields
\begin{equation}
\label{ansatz2}
[\mathsf\Omega(t)\boldsymbol\cdot\mathsf R(t)\boldsymbol\cdot(\boldsymbol r-\boldsymbol r_c(t))-\boldsymbol v(t)]\boldsymbol\cdot\boldsymbol C'(\tilde{\boldsymbol r})
\equiv [\mathsf \Omega(t)\boldsymbol\cdot\tilde{\boldsymbol r}-\boldsymbol v(t)]\boldsymbol\cdot\boldsymbol C'(\tilde{\boldsymbol r})=\mathcal G\{C(r')\}(\tilde{\boldsymbol r}).
\end{equation}
Using that the components of the gradient $\boldsymbol C'(\tilde{\boldsymbol r})$ are linearly independent functions of $\tilde{\boldsymbol r}$, we get from (\ref{ansatz2}) that $\mathsf \Omega(t)$ and $\boldsymbol v(t)$ {are actually constant in time}.
From this observation we write $\mathsf \Omega(t)=\mathsf \Omega^0$, $\boldsymbol v(t)=\boldsymbol v^0$.
It follows from (\ref{Omega}) that $\mathsf\Omega^0$ is an antisymmetric matrix.
It can thus be written as $\Omega^0_{ij}=\epsilon_{ijk}\omega^0_k$, where $\epsilon$ is the Levi-Civita symbol and $\boldsymbol\omega^0$ is some vector.

An intuitive explanation of the above derivation can be given as follows:
$C(\tilde{\boldsymbol r})$ corresponds to the concentration field in the frame of reference comoving and corotating with the concentration field.
Since this field is time-constant, all functions associated with it should also be constant in the comoving and corotating frame.
In particular, the velocity and the angular velocity of the fixed laboratory frame should also be constant when measured and expressed in the comoving and corotating frame.
The vectors $-\boldsymbol v^0$ and $-\boldsymbol\omega^0$ represent these constants.

The vector $\boldsymbol r_c(t)$ defines the position of the particle, while the matrix $\mathsf R(t)$ defines its orientation (a spherical particle does not have intrinsic orientation; it is the angular dependence of the concentration field that sets the orientation).
Finding the trajectory requires solving for $\boldsymbol r_c(t)$.
Integrating (\ref{Omega}) yields
\begin{equation}
\label{Rt}
\mathsf R(t)=\exp(\mathsf\Omega^0t)\boldsymbol\cdot\mathsf R(0) .
\end{equation}
For simplicity, we choose the axes in the corotational frame to coincide with the axes of the laboratory frame at $t=0$.
The matrix $\mathsf R(0)$ is the identity matrix in this case.
We choose the $z$ axis as the direction of $\boldsymbol\omega$ at $t=0$.
The matrix $\mathsf\Omega$ then has only two non-zero components for this choice of axes: $\Omega_{xy}=-\Omega_{yx}=\omega^0$.
Its exponent then reads
\begin{equation}
\label{exp3D}
\exp(\mathsf\Omega^0t)=
	\begin{pmatrix}
		\cos\omega^0 t & \sin\omega^0 t & 0 \\
		-\sin\omega^0 t & \cos\omega^0 t & 0 \\
		0 & 0 & 1
	\end{pmatrix}
,
\end{equation}
which corresponds to a usual rotation matrix.
We can now solve for the propulsion velocity
\begin{equation}
\label{vc}
\dot {\boldsymbol r}_c(t)=\mathsf R(t)^{-1}\boldsymbol\cdot \boldsymbol v^0=v^0_x(\boldsymbol e^x \cos \omega^0 t+\boldsymbol e^y \sin\omega^0 t)+v^0_z\boldsymbol e^z,
\end{equation}
where we have chosen the $y$ axis at $t=0$ in such a way that $v^0_y=0$.
$\boldsymbol e^x$, $\boldsymbol e^y$, $\boldsymbol e^z$ are unit vectors along the corresponding directions.

Let us now analyze the possible dynamics defined by Eq. (\ref{vc}) and the resulting trajectories.
The simplest case corresponds to $v^0_x=v^0_z=0$, whence $\dot {\boldsymbol r}_c=0$.
The particle is stationary in this case.
The next possibility is $v^0\ne 0$ and $\omega^0=0$.
The velocity (\ref{vc}) is constant in time in this case.
The trajectory of the particle is a straight line.
Next, we consider the case $v^0_x\ne 0$, $\omega^0\ne 0$ and $v^0_z=0$.
The velocity is time-dependent in this case.
It always lies in the $(x,y)$ plane and rotates about the $z$ axis with a constant angular velocity $\omega^0$.
The corresponding trajectory is a circle in the $(x,y)$ plane.
The particle traces a full circle in time $2\pi/\omega^0$, during which time it travels a distance of $2\pi v^0/\omega^0$.
This corresponds to a circle of radius $v^0/\omega^0$.
Finally, if $v^0_x\ne 0$, $\omega^0\ne 0$, and $v^0_z\ne 0$, then the velocity in the laboratory frame has a rotating component in the $(x,y)$ plane and a constant component along the $z$ axis.
Integrating the velocity in time shows that the trajectory corresponds to a helix, whose axis is parallel to $z$ axis.
The period of the rotating component is $2\pi/\omega^0$, during which time the particle projection on the $(x,y)$ plane travels a distance of $2\pi v^0_x/\omega^0$.
This implies that the radius of the helix is $v^0_x/\omega^0$.
The distance traveled along the $z$ axis during one period is $2\pi v^0_z/\omega^0$, which sets the pitch of the helix.
The following expressions of the helix radius $a$ and the pitch $2\pi b$ can be used for an arbitrary choice of the coordinates:
\begin{equation}
\label{helix}
a=\frac{|\boldsymbol v^0\boldsymbol\times\boldsymbol\omega^0|}{(\omega^0)^2},\,\,\,2\pi b=2\pi\frac{\boldsymbol v^0\boldsymbol\cdot\boldsymbol\omega^0}{(\omega^0)^2}.
\end{equation}

The trajectory type is related to the symmetry of the concentration field $C$.
Namely, if any transformation of the space keeps $C$ invariant, the vectors $\boldsymbol v^0$ and $\boldsymbol \omega^0$ should also be unchanged by this transformation.
It is important to note that vector $\boldsymbol v^0$ is polar, while the vector $\boldsymbol \omega^0$ is axial, that is the former changes sign on inversion of the space and the latter is unaffected by it.
Consider now the 4 trajectory types:
In non-motile state both vectors are zero, so the symmetry of the $(\boldsymbol v^0,\boldsymbol \omega^0)$ tuple is the full $O(3)$ group.
Straight trajectory corresponds to $\omega^0=0$, so the symmetry group is $C_{\infty v}$, which corresponds to arbitrary rotations about the $\boldsymbol v^0$ direction and also reflection with respect to any plane containing $\boldsymbol v^0$.
The circular trajectory corresponds to a single mirror symmetry (group $C_s$) with respect to the plane containing $\boldsymbol v^0$ and orthogonal to $\boldsymbol \omega^0$ (as it is an axial vector).
This mirror plane defines the plane in which the circle is drawn.
Finally, the vectors $\boldsymbol v^0$ and $\boldsymbol \omega^0$ are not orthogonal for helical trajectories, which leaves the symmetry group of the system trivial.

Phoretic particles in lower dimensions can exhibit a subset of the 4 trajectory types described above:
Stationary particle, straight and circular motions in 2D, or stationary particle and straight motion in 1D.

\section{Emergence of self-congruent solutions}
The question  arises of whether self-congruent solutions are just a mathematical curiosity, or whether they emerge naturally in a given motility model. It will appear here and in subsequent sections that self-congruent solutions are the rule. In this Section we analyze 
the emergence of motile self-congruent solutions from a stationary state and the bifurcations which mark the transitions from one trajectory type to another. Our starting point is the evolution equation (\ref{functional}).
We first consider  the simplest case of 1D phoretic system and then show how the dynamics in two or more dimensions can be reduced to the 1D case.
\subsection{1D case: segment particle on a line}
\subsubsection{Stationary solution}
The 1D case corresponds to a distribution of concentration $c(x)$ along the line $x\in(-\infty,\infty)$.
A related discussion can be found in \cite{Ophaus2018} in application to a particular equation governing a phase field crystal in 1D. 
We take the particle as a segment which runs from $x=x_c-1$ to $x=x_c+1$, where $x_c$ is the position of the particle.
We place initially the particle in the origin, such that $x_c=0$ at $t=0$.
The functional dependence of the time evolution of the concentration field for its given distribution is given by Eq. (\ref{functional}), where $c$ is a function of the only variable $x$.
There are two symmetries that the dependence (\ref{functional}) must satisfy in 1D:
the translational invariance
\begin{equation}
\label{translational}
\mathcal G\{c(x'+x_0)\}(x)=\mathcal G\{c(x')\}(x+x_0).
\end{equation}
and the change of sign of $x$:
\begin{equation}
\label{reflection}
\mathcal G\{c(-x')\}(x)=\mathcal G\{c(x')\}(-x).
\end{equation}

We first take the limit of low activity $Pe$, which corresponds to a ground state $c_0(x)$ of the concentration distribution.
It is natural to assume that this distribution is symmetric $c_0(-x)=c_0(x)$ and thus by symmetry the particle is not motile.
By definition (stationary solution), $\mathcal G\{c_0(x')\}=0$.
The next step is to analyze the linear stability of this solution.
We introduce a small perturbation of the concentration field $\varepsilon\delta c(x)$, where $\varepsilon$ is a small parameter.
Taking $c(x)=c_0(x)+\varepsilon\delta c(x)$, we can write
\begin{equation}
\label{Gexpansion}
\mathcal G\{c_0(x')+\varepsilon\delta c(x')\}(x)=\varepsilon \mathcal G_1\{\delta c(x')\}(x)+\varepsilon^2 \mathcal G_2\{\delta c(x'),\delta c(x'')\}(x)+\varepsilon^3\mathcal G_3\{\delta c(x'),\delta c(x''),\delta c(x''')\}(x)+O(\varepsilon^4).
\end{equation}
Here $\mathcal G_1$, $\mathcal G_2$, and $\mathcal G_3$ are linear, bilinear, and trilinear operators, respectively.
The linear stability of the solution $c_0(x)$ is thus related to the eigenvalues of the linear operator $\mathcal G_1$.
Further below, we use $\lambda_i$ and $f_i(x)$ to denote the eigenvalues and the corresponding eigenfunctions of the linear operator $\mathcal G_1$.
Here the index $i$ is supposed to run from 0 to $\infty$.
{It is important for our analysis that the set of eigenvalues be discrete.
This is usually the case for systems of finite size, as happens for autophoretic particles in a finite fluid domain or for myosin-induced motility in which the myosin is localized within the cell cortex.
}

An important property of the eigenfunctions $f_i(x)$ is that they are either symmetric or antisymmetric with respect to the transformation $x\rightarrow -x$.
This is a consequence of the symmetry of the operator $\mathcal G_1$ and the solution $c_0(x)$.
Indeed, the function $f_i(-x)$ is an eigenfunction of the operator $\mathcal G_1$ with the same eigenvalue $\lambda_i$.
This means that the two functions $f_i(x)$ and $f_i(-x)$ are linearly dependent.
The proportionality  coefficient can only be 1 or -1. Indeed if ${\cal P}$ is the inversion operator. (which commutes with $\mathcal G_1$, and thus has common eigenmodes), it obeys ${\cal P}^2=I$, meaning that its eigenvalues obey  $\lambda^2=1$ ($\lambda=\pm 1$). Consequently, since by definition ${\cal P} f(x)=f(-x)=\lambda f(x)$, and because $\lambda=\pm 1$, we get for the two eigenstates  $f_1(-x)=f_1(x)$ and $f_2(-x)=-f_2(x)$

One eigenvalue of $\mathcal G_1$ is special and is equal to zero for any $Pe$.
This eigenvalue corresponds to the homogeneous translation of the concentration field.
Indeed, substituting $c(x)=c_0(x+\varepsilon)=c_0(x)+\varepsilon\partial_xc_0(x)+O(\varepsilon^2)$  (meaning $\delta c= \partial_xc_0(x)$) in Eq. (\ref{Gexpansion}), and using $\mathcal G\{c_0(x+\varepsilon)\}=\mathcal G\{c_0(x)\}=0$,  we get
\begin{equation}
\label{zeroeigen}
\mathcal G_1\{\partial_{x'} c_0(x')\}(x)=0.
\end{equation}
That is, the eigenfunction $\partial_x c_0(x)$ corresponds to the eigenvalue 0.
Further below, we use index 0 for this eigenvalue ($\lambda_0=0,$ $f_0(x)=\partial_x c_0(x)$).

\subsubsection{Critical activity and the importance of non self-adjoint operator $\mathcal G_1$}

The solution $c_0(x)$ is linearly stable provided all eigenvalues other than 0 have negative real parts.
We now analyze the change of the dynamics of the system as one of the eigenvalues goes from negative to positive while the real parts of all the other eigenvalues remain negative.
Suppose we have an eigenvalue $\lambda_1$ and a critical activity $Pe_1$ such that $\lambda_1(Pe_1)=0$.
We also consider the case when the eigenfunction $f_1$ is antisymmetric and not symmetric, so that the symmetry of the system changes at $Pe=Pe_1$.
The goal of the following discussion is to show that the dynamics of the particle undergoes a pitchfork bifurcation at $Pe=Pe_1$ and that if the bifurcation is supercritical the concentration field for $Pe>Pe_1$ corresponds to a self-congruent solution with translational velocity $v^0\propto(Pe-Pe_1)^{1/2}$.

The operator $\mathcal G_1$ has two eigenvalues equal to 0 for $Pe=Pe_1$.
Only if the operator $\mathcal G_1$ is self-adjoint, can it be fully diagonalized for $Pe=Pe_1$.
This would be the case, for example,  if the problem were variational (global thermodynamical equilibrium).
In active systems the evolution equation  (\ref{functional}) is non-variational and the linearized operator is generically not self-adjoint.
This will have important consequences as seen below.
Non self-adjoint operator $\mathcal G_1$ with a degenerate eigenvalue is called defective.
It can not be fully diagonalized.

In the case of non self-adjoint operator, with a doubly degenerate eigenvalue ($\lambda=0$), we use the Jordan normal form to analyze the linear dynamics of the system:
Equation (\ref{zeroeigen}) remains valid for $Pe=Pe_1$ because it represents a symmetry of the problem.
In the Jordan spirit, if $f_0(x)$ is an eigenfunction associated to a zero eigenvalue (and it is so as we have shown above), then another function $f_1(x)$  satisfies $\mathcal G_1\{f_1(x')\}(x)=f_0(x)\equiv\partial_xc_0(x)$ for $Pe=Pe_1$.
The function $f_1(x)$ is called a generalized eigenfunction.
Substituting $c(x)=c_0(x)+\varepsilon f_1(x)$ in (\ref{Gexpansion}), we get 
\begin{equation}
\label{advection}
\dot c(x)=\varepsilon \partial_xc_0(x)+O(\varepsilon^2).
\end{equation}
The form of Eq. (\ref{advection}) is that of an advection equation, the general  solution of  which does not depend on $x$ and $t$ separately, but on $x-v_0t$ ($v_0$ is a constant).
To leading order the concentration field reads
\begin{equation}
\label{move}
c(x,t)=c_0(x-v_0t)+\varepsilon f_1(x-v_0t)+O(\varepsilon^2).
\end{equation}
The concentration evolution equation then reads
\begin{equation}
\label{movesubs}
-v^0\partial_xc_0(x-v^0t)=\varepsilon\partial_xc_0(x-v^0t)+O(\varepsilon^2),
\end{equation}
whence $v^0=-\varepsilon$.
This confirms that for $Pe=Pe_1$ the solution (\ref{move}) is marginally stable and would move in the negative $x$ direction with velocity $\varepsilon$.
Thus the first type of motile self-congruent solutions (motion at constant velocity) emerges here naturally. Note that if the linear operator were self-adjoint, then we would have $\mathcal G_1 (f_1)=0$ at bifurcation point, and the right hand side of (\ref{movesubs}) would be zero, and so does the swimming speed. Note also that if the nonlinear operator $\mathcal G$ (Eq. (\ref{functional}) is a functional derivative of some functional (implying automatically that the linear operator is self-adjoint) then there is no swimming. A general proof is given in Appendix \ref{app1}. 
\subsubsection{Above critical activity}
The next step is to solve the problem for positive values of $Pe-Pe_1$ in order to find the value of $\varepsilon$ as a function of $Pe$.
The eigenvalues $\lambda_0$ and $\lambda_1$ of the operator $\mathcal G_1$ are distinct for $Pe\ne Pe_1$ and it can thus be fully diagonalized.
However, the eigenvectors $f_0$ and $f_1$ are almost identical (up to a scale factor) as $Pe$ tends to $Pe_1$.
A basis containing these vectors is therefore not particularly suitable for representing the solution close to the bifurcation point.
Instead, we use the generalized eigenfunctions of the operator $\mathcal G_1$ evaluated at $Pe=Pe_1$ to parametrize the solution.
These functions are thus independent of $Pe$, while the operator $\mathcal G_1$ undergoes a regular perturbation at the bifurcation point.
This choice of the basis functions makes the calculations straightforward but the expressions remain quite involved.
We therefore give here only the main idea of the derivation while the details are given in Appendix \ref{Appendix}.

The concentration field is written as
\begin{equation}
\label{comoving}
c(x,t)=c_0(x-x^0(t))+\varepsilon(t) f_1(x-x^0(t))+\sum\limits_{i=2}^\infty \delta c_i(t)f_i(x-x^0(t)),
\end{equation}
in the comoving frame.
Here $x^0(t)$ sets the position of the particle.
The expansion (\ref{comoving}) represents the concentration field as a sum of the non-motile solution and a perturbation, which is expanded in the basis of the generalized eigenfunctions of the operator $\mathcal G_1$.
Note that the function $f_0(x)$ is missing from the expansion (\ref{comoving}).
Its role is taken by the variable position $x^0$.
This allows us to convert the evolution of the continuous concentration field into a discrete system of ordinary differential equations.
Namely, $\mathcal G_1\{c\}$ is expanded into the basis defined by the functions $f_i$, which gives the time derivatives of $x_0(t)$, $\varepsilon(t)$, and $\delta c_i(t)$.
Next we derive a closed evolution equation for $\varepsilon(t)$.
This can be done by so-called adiabatic elimination of the rapidly decaying modes in the concentration perturbation (the companion Letter \cite{misbahLetter} gives a practical example of this technique).
Indeed, the phoretic system has two distinct time scales close to the critical activity:
One corresponds to the evolution of $\varepsilon$, the growth rate of which is proportional to $Pe-Pe_1$ and thus can be arbitrary small close enough to the transition point.
The other time scale is defined by the smallest of $|\lambda_i|$ for all $i>1$, which sets the decay rate of the stable modes.
The latter time scale remains finite at the transition point.
The adiabatic elimination proceeds in the following way:
\begin{enumerate}
\item We fix a value of $\varepsilon$.
\item We solve the time evolution equations for $\delta c_i$ with this value of $\varepsilon$.
\item We obtain the steady-state values to which $\delta c_i(t)$ relax after an initial transient.
\item We call these saturation values as $\delta c_i^0(\varepsilon)$.
\item We substitute the values $\delta c_i^0(\varepsilon)$ in equations for $\dot\varepsilon(t)$ and $\dot x^0(t)$.
\end{enumerate}
The above procedure converges to fixed values of $\delta c_i^0(\varepsilon)$ for sufficiently small $\varepsilon$ and $Pe-Pe_1$ because all modes $f_i(x)$ for $i>1$ are stable at $\varepsilon=0$ and $Pe=Pe_1$.
The values $\delta c_i^0(\varepsilon(t))$ agree with the actual solution $\delta c_i(t)$ to the leading order thanks to the separation of time scales.
In particular, $\delta c_i^0(\varepsilon(t))$ is equal to $\delta c_i(t)$ exactly for steady-state solutions.

Once the adiabatic elimination is applied, the time evolution equation for $\varepsilon(t)$ can be expanded as:
\begin{equation}
\label{epsilondot}
\dot\varepsilon(t)=a_1\varepsilon(t)+a_3\varepsilon(t)^3+O(\varepsilon(t)^5).
\end{equation}
The coefficients for even powers in expansion (\ref{epsilondot}) vanish because the function $f_1(x)$ is odd, so that the concentration field must be invariant under transformation $(x,\varepsilon,t)\rightarrow(-x,-\varepsilon,t)$, which in the case of Eq. (\ref{epsilondot}) reduces to $\varepsilon\rightarrow-\varepsilon$ invariance.
The coefficient $a_1$ is proportional to $Pe-Pe_1$ close to the bifurcation point.
If the coefficient $a_3$ is negative, Eq. (\ref{epsilondot}) has two stable fixed points $\varepsilon^0=\pm(-a_1/a_3)^{1/2}\propto(Pe-Pe_1)^{1/2}$.
This defines a supercritical pitchfork bifurcation.
Equation (\ref{movesubs}) remains valid to the leading order even for $Pe>Pe_1$, which implies $\dot x^0=-\varepsilon+o(\varepsilon)$.
The presented analysis shows that the concentration evolution equation has a stable self-congruent solution of form (\ref{move}) with velocity proportional to $(Pe-Pe_1)^{1/2}$. Appendix  \ref{Appendix} provides an explicit calculation.

\subsection{2D case: circular particle in a plane}

A generic example of the bifurcation dynamics in a 2D autophoretic system is considered in the companion Letter \cite{misbahLetter}.
Among the bifurcations presented in that work, two correspond to transitions to self-congruent dynamics:
The first one marks transition from a stationary isotropic concentration field around the particle to a polarized concentration field and straight motion of the particle.
The concentration field retains a mirror symmetry after this bifurcation.
The second bifurcation marks transition from straight to circular trajectories.
The concentration field loses its mirror symmetry at this bifurcation.
Here we show how these observations can be explained in the general framework of the self-congruent solutions.

\subsubsection{Primary bifurcation}

For the 2D case we consider a concentration field $c(x,y)$ defined on a plane.
It is convenient to introduce a comoving polar coordinate system by associating the center of the particle with the origin.
For low enough $Pe$, the concentration field is rotationally symmetric and thus is a function of distance from the center $r$, $c_0(x,y)=c_0(r)$.
This corresponds to the non-motile state.
As in the 1D case, we analyze the linear stability of this solution.
We define the operator $\mathcal G_{1}\{c(x',y')\}(x,y)$ in the same way as for the 1D case.
For each eigenvalue $\lambda_i$ of $\mathcal G_{1}\{c(x',y')\}(x,y)$, the corresponding eigenfunctions $f_{i,j}(x,y)$ introduce an irreducible representation of $O(2)$, the rotation and reflection symmetry group of the unperturbed solution.
The subscript $j$ is used to index the eigenfunctions corresponding to an eigenvalue $\lambda_i$.
$O(2)$ is the group of all rotations about the origin and reflections in straight lines passing through the origin.
As is generally known for the representations of $O(2)$, an eigenvalue $\lambda_i$ corresponds either to a single rotationally symmetric eigenfunction 
\begin{equation}
\label{eigenmode2D1}
f_{i,1}(x,y)=f_i(r),
\end{equation}
or to a pair of functions 
\begin{equation}
\label{eigenmode2D2}
f_{i,1}(x,y)=f_i(r)\cos m_i\phi,\,\,\,f_{i,2}(x,y)=f_i(r)\sin m_i\phi,
\end{equation}
where $\phi$ is the polar angle.
Here $m_i$ is a natural number that defines the angular dependence of the eigenfunctions $f_{i,j}(x,y)$ for given $i$.
An explicit derivation of the form (\ref{eigenmode2D1}), (\ref{eigenmode2D2}) of the eigenfunctions is classically known for the Laplace operator $\nabla^2$ in polar coordinates.

As in  the 1D case, the derivatives $\partial_x c_0(x,y)$, $\partial_y c_0(x,y)$ are eigenfunctions of $\mathcal G_{1}$ with the corresponding eigenvalue equal to zero.
Recall that these are translation modes.
Indeed, displacing the particle and the concentration field along $x$ or along $y$ by constant values corresponds to neutral modes.  
As the activity is increased, one of the eigenvalues of the linear stability operator becomes positive at a critical activity $Pe_1$ and the rotationally symmetric solution becomes unstable.
If this eigenvalue corresponds to eigenfunctions (\ref{eigenmode2D2}) with $m_i=1$, the resulting solution is motile.
The first unstable mode was indeed observed to have $m_i=1$ in many models of autophoretic particles and of mammalian cell swimming \cite{rednikov1994drop,golovin1989change, MLB13,morozov2019nonlinear,hawkins2011spontaneous,Voituriez2016,Farutin2019}.

The emerging dynamics corresponds to a self-congruent solution moving with a constant velocity along a straight line.
The velocity scales as $(Pe-Pe_1)^{1/2}$.
This can be demonstrated in the same way as for the 1D case.
First, we assume that the concentration field is symmetric with respect to the reflection $y\rightarrow -y$.
With this assumption, the perturbation of the concentration field $\delta c(x,y)$ does not contain the modes $\partial_y c_0(x,y)$ and $f_i(r)\sin m\phi$ because they are antisymmetric with respect to the $y\rightarrow -y$ reflection.
Extending the ansatz (\ref{comoving}) to 2D then yields eq. (\ref{epsilondot}) as in the 1D case.
Thus we obtain that the $y\rightarrow -y$ symmetric concentraton field looses its $x\rightarrow -x$ symmetry at $Pe=Pe_1$ and the swimmer starts to move along the $x$ direction.
A question remains whether this motion is stable with respect to perturbations that do not respect the $y\rightarrow -y$ symmetry.
As we show below, this solution is marginally stable for $Pe$ sufficiently close to $Pe_1$: 
After a perturbation without $y\rightarrow -y$ symmetry is applied, the solution relaxes to the same motion but with some shift and rotation compared to the original one.

\subsubsection{Secondary bifurcation}

As demonstrated in the companion Letter \cite{misbahLetter}, the straight motion can become unstable resulting in a circular trajectory of the particle.
We analyze this secondary bifurcation by reducing the dimension of the problem.
First, we rewrite the evolution equation (\ref{functional}) in the reference frame comoving with the particle.
Next, we transform the full 2D problem into a simplified 1D problem in the $\phi$ space for a given value of $r$.
This would allow us to use the results of the previous section to analyze the instability.

The perturbation dynamics of the motile solution $c_1(x,y)\equiv c_1(r,\phi)$ possesses 3 soft modes (i.e. neutral modes):
Two are related to translations, while the third one is related to rotation.
The last mode relates to the fact that $c_1(r,\phi+\phi_0)$ is also a solution.
The other perturbation modes (other than soft ones) of the $c_1(r,\phi)$ solution should correspond to negative eigenvalues for low enough values of the activity, which ensures the stability of the straight motion, as discussed below.
As the activity increases, one of the perturbation modes can become unstable.
The resulting dynamics can be deduced from applying the analysis developed for the 1D case to the distribution $c(r,\phi)$ as a function of $\phi$ for a fixed value of $r$.
For this reason we only list schematically the main steps of the analysis. The companion Letter \cite{misbahLetter} provides an explicit calculation (in the Letter $\phi$ is denoted as $s$).
The concentration field of a particle moving along the $x$ axis possesses the $\phi\rightarrow-\phi$ (equivalently, $y\rightarrow-y$) symmetry, as discussed above.
Like in 1D case, this implies that all eigenfunctions of the linear perturbation operator are either symmetric or antisymmetric with respect to the change of the sign of $\phi$ (equivalently, $y$).
The secondary symmetry breaking bifurcation (loss of $y\rightarrow -y$ symmetry) occurs if the perturbation mode that loses stability is antisymmetric in $y$.
Above the critical value $Pe_2$, this mode grows exponentially until it saturates with amplitude that scales as $(Pe-Pe_2)^{1/2}$ similarly to the 1D case, provided the bifurcation is supercritical.
Once the amplitude of the unstable mode reaches the steady-state value, the system follows self-congruent dynamics.
Indeed, the loss of $\phi\rightarrow-\phi$ symmetry entails a self-congruent motion of the concentration field in the $\phi$ domain, which corresponds to the rotational soft mode of the solution $c_1(r,\phi)$.
The concentration maximum drifts along the particle periphery at a constant speed.
This implies that the particle moves along a circular trajectory for $Pe>Pe_2$, as shown in Section \ref{sect:SelfCongruent}.
The angular drift velocity  scales as $\omega^0\propto(Pe-Pe_2)^{1/2}$, exactly as a drift along $x$-direction in 1D.
The steady-state value of translational velocity is continuous at the transition point.
The radius of the circle scales as $(Pe-Pe_2)^{-1/2}$ according to the analysis in Section \ref{sect:SelfCongruent} and the calculation of the companion Letter\cite{misbahLetter}.

\subsubsection{Perturbation spectrum}

\begin{figure}
\begin{center}
\includegraphics[width=0.9\columnwidth]{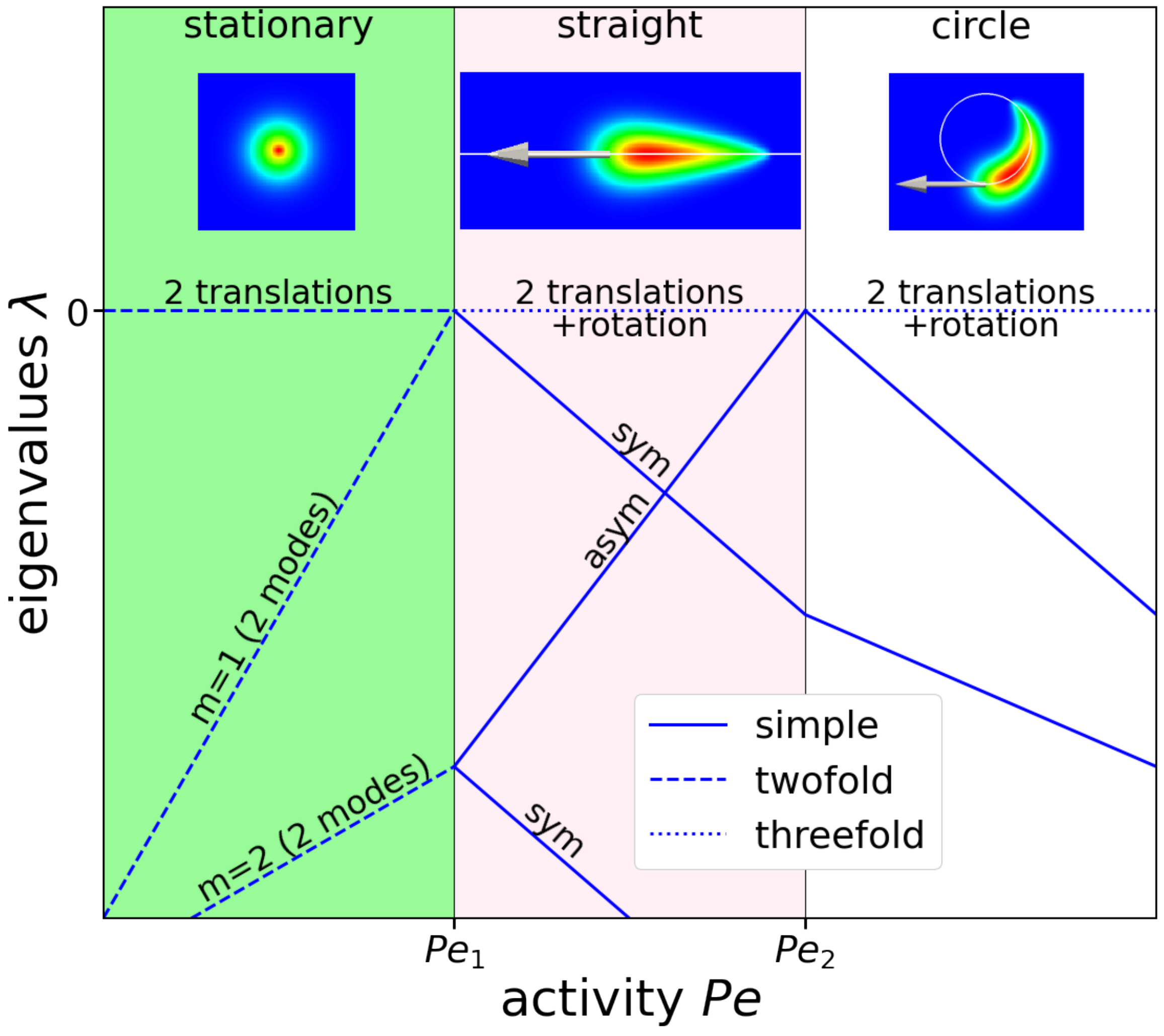}
\caption{\label{eigenvalues}Schematic view of the growth rate of the most unstable modes as a function of the activity. The modes symmetric and antisymmetric with respect to the $y\rightarrow -y$ reflection are marked as sym and asym, respectively.
Insets show the characteristic dynamics of the particle in each phase. Color code shows the concentration distribution. Arrows show the velocity of motile particles. White curves show the trajectories.
}
\end{center}
\end{figure}

In order do provide further insight into the nature of the two bifurcations that separate the stationary phase (the non-motile solution) and the two motile phases from (the straight and circular ones) each other, we plot schematically the growth rate of the least stable eigenmodes of the stable  solution (as a function of $Pe$ in Fig. \ref{eigenvalues}).
We first analyze the behavior of eigenvalues in the stationary phase (the non motile one):
Each eigenvalue (except for those that correspond to rotationally symmetric modes; not shown in Fig. \ref{eigenvalues}) is twofold degenerate, where the two eigenfunctions differ by $\sin$/$\cos$ in the angular dependence (\ref{eigenmode2D2}).
The two eigenvalues corresponding to the translation modes are exactly 0.
Another branch (which is doubly degenerate, and marked $m=1$ in Fig. \ref{eigenvalues}) corresponds to the mode that becomes unstable  at $Pe=Pe_1$.
This branch has a negative value but increases with $Pe$.
It reaches 0 for $Pe=Pe_1$ which corresponds to the primary bifurcation.
There are thus four eigenvalues equal to 0 at $Pe=Pe_1$ (two translation modes and the twofold degenerate --sin/cos -- branch that becomes unstable at $Pe=Pe_1$).

We now discuss how these four branches are continued into the motile phase:
First, we note that the growth rate of small perturbations needs to be measured in the reference frame comoving with the unperturbed solution for the motile phases.
Second, because the transition to the motile phase is continuous, the growth rate of eigenmodes evolves continuously across the transition point $Pe=Pe_1$.
Third, we take the $y\rightarrow -y$ symmetric solution discussed above as the unperturbed motile solution.
Our goal here is to show that the growth rate is negative for all possible perturbations of this solution, at least for some range of $Pe$, which would prove its stability.
Since the rotational symmetry is broken in the motile phase, the eigenmodes can not be written as eqs. (\ref{eigenmode2D1}) and (\ref{eigenmode2D2}) (which are valid only for perturbations of the fully symmetric --non motile-- solution).
Nevertheless, because the unperturbed solution possesses the $y\rightarrow -y$ symmetry, the perturbation eigenmodes can be chosen to be either symmetric or antisymmetric with respect to the $y\rightarrow -y$ reflection, as was discussed in the 1D case (where we saw that eigenmodes are either symmetric or antisymmetric with respect tp $x$).
The eigenvalue branches are, in general, not degenerate in the motile phase (except for the neutral modes).
This is because the $x$ and $y$ directions are not equivalent for a polarized concentration field.
This means that each twofold degenerate branch (other than the translation branch) splits at $Pe=Pe_1$ into two simple branches.
One of those branches is symmetric with respect to the $y\rightarrow -y$ reflection (continuation of the $\cos m\phi$ eigenmode in (\ref{eigenmode2D2})), while the other one is antisymmetric (continuation of the $\sin m\phi$ eigenmode in (\ref{eigenmode2D2})).

There are 3 neutral eigenmodes (growth rate equal to 0) for $Pe>Pe_1$: two translations and 1 rotation of the concentration field.
Of these, one translation mode (along $x$) is symmetric with respect to $y\rightarrow -y$ reflection.
The other translation mode and the rotation mode are antisymmetric with respect to $y\rightarrow -y$ reflection.
The fourth mode equal to 0 at $Pe=Pe_1$ is symmetric with respect to the $y\rightarrow -y$ reflection.
It corresponds to the linear stability of the non-zero solution of eq. (\ref{epsilondot}) (i.e. the mode corresponding to motility along $x$ is stable for $Pe<Pe_2$).
Its growth rate is thus negative for $Pe>Pe_1$ due to the supercritical nature of the bifurcation, as discussed in the 1D case.
We thus conclude that all 4 eigenvalue branches emerging from 0 at $Pe=Pe_1$ remain non-positive for $Pe>Pe_1$.
This confirms that the $y\rightarrow -y$ symmetric motile solution is marginally stable in some region above $Pe_1$.

Consider now the transition from straight to circular trajectories.
This transition happens at $Pe=Pe_2$ and requires a mode which is antisymmetric under the $y\rightarrow -y$ reflection to become unstable, as discussed above.
This mode can not be a continuation of the $m=1$ mode in Fig. \label{eigenvalues} because the antisymmetric $m=1$ mode continues as the neutral rotation mode in the motile phases.
We therefore conclude that the instability occurs due to another mode, the growth rate of which is negative for the stationary and straight phases but increases reaching zero at $Pe=Pe_2$.
It is shown in the companion Letter\cite{misbahLetter} that including the first two harmonics ($m_i=1,2$ in eq. (\ref{eigenmode2D2}) in the concentration field is sufficient to observe both the primary and the secondary instability.
For this reason, we have marked the mode that loses stability at $Pe=Pe_2$ as $m=2$ mode, which is a natural choice but not a general rule.

We plot the growth rate of the perturbation eigenmodes for the circular trajectory for $Pe>Pe_2$.
Since this solution does not satisfy any spatial symmetry, neither do the corresponding perturbation modes.
The continuation of the antisymmetric $m=2$ mode has a negative growth rate for $Pe>Pe_2$ as is expected for a linearly stable state.

\begin{figure*}[ht]
\centering
\begin{tabular}{l|c|c|c|c}
column tag & A & B & C & D \\
\hline
concentration in 2D & \includegraphics[width=0.15\textwidth]{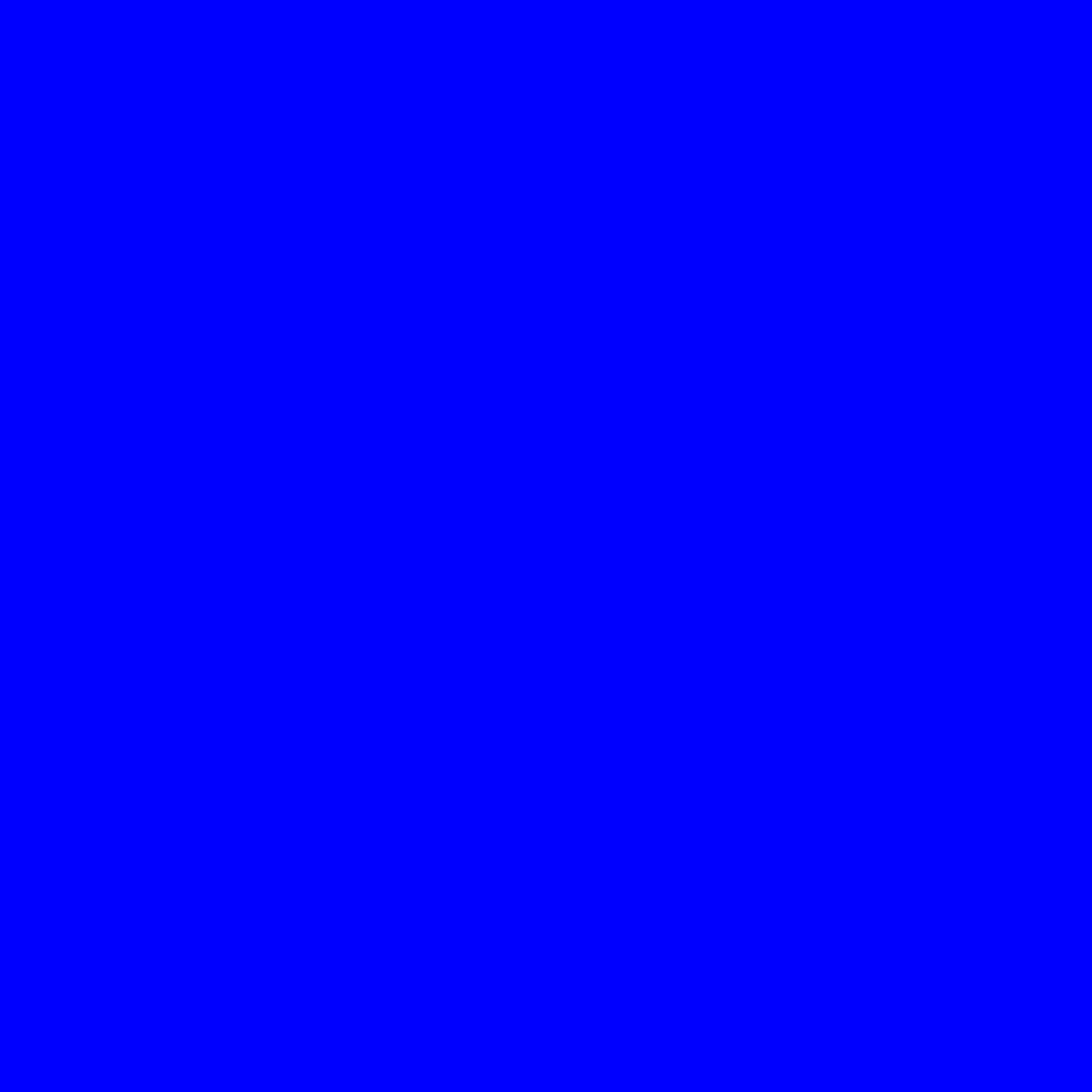} &\includegraphics[width=0.15\textwidth]{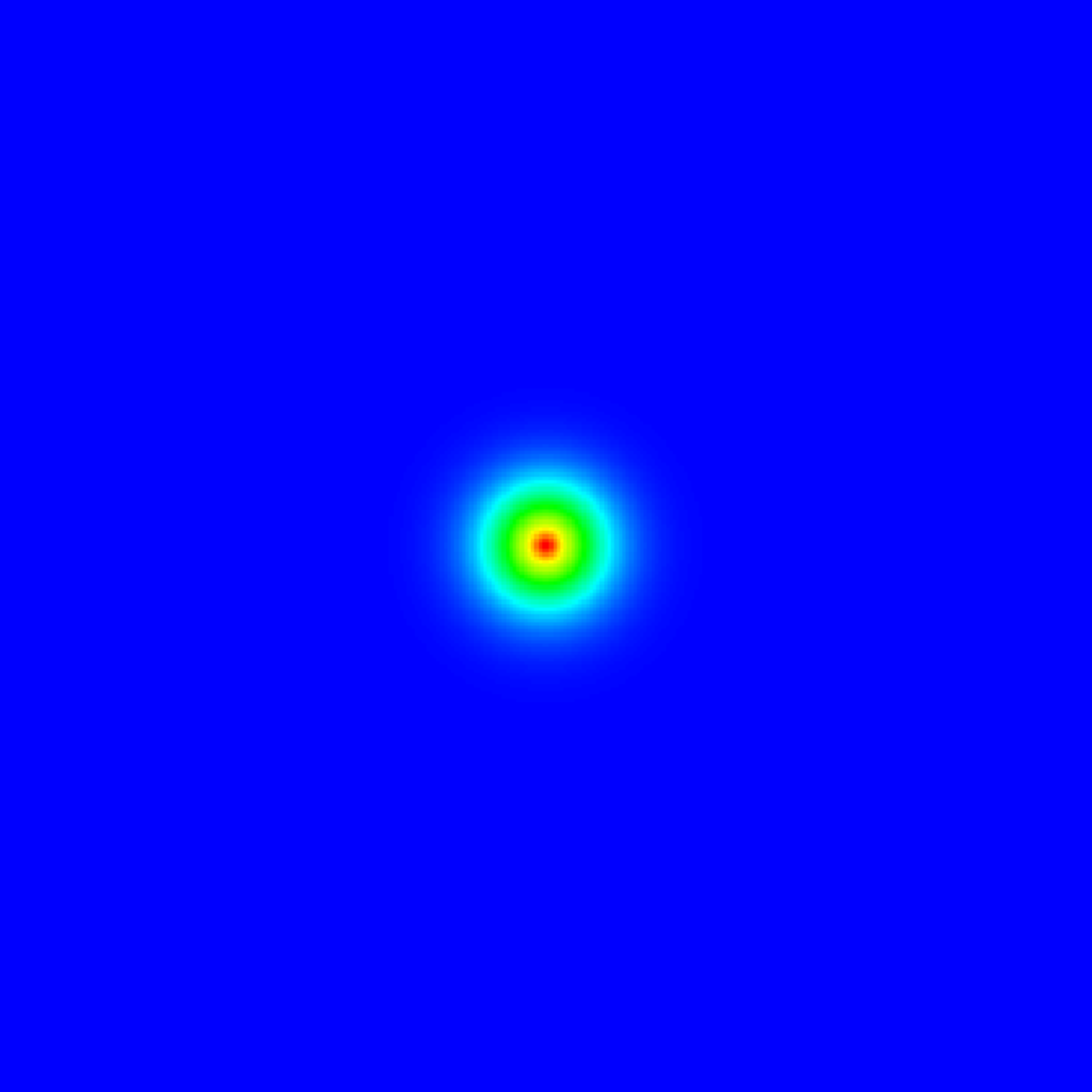} &\includegraphics[width=0.15\textwidth]{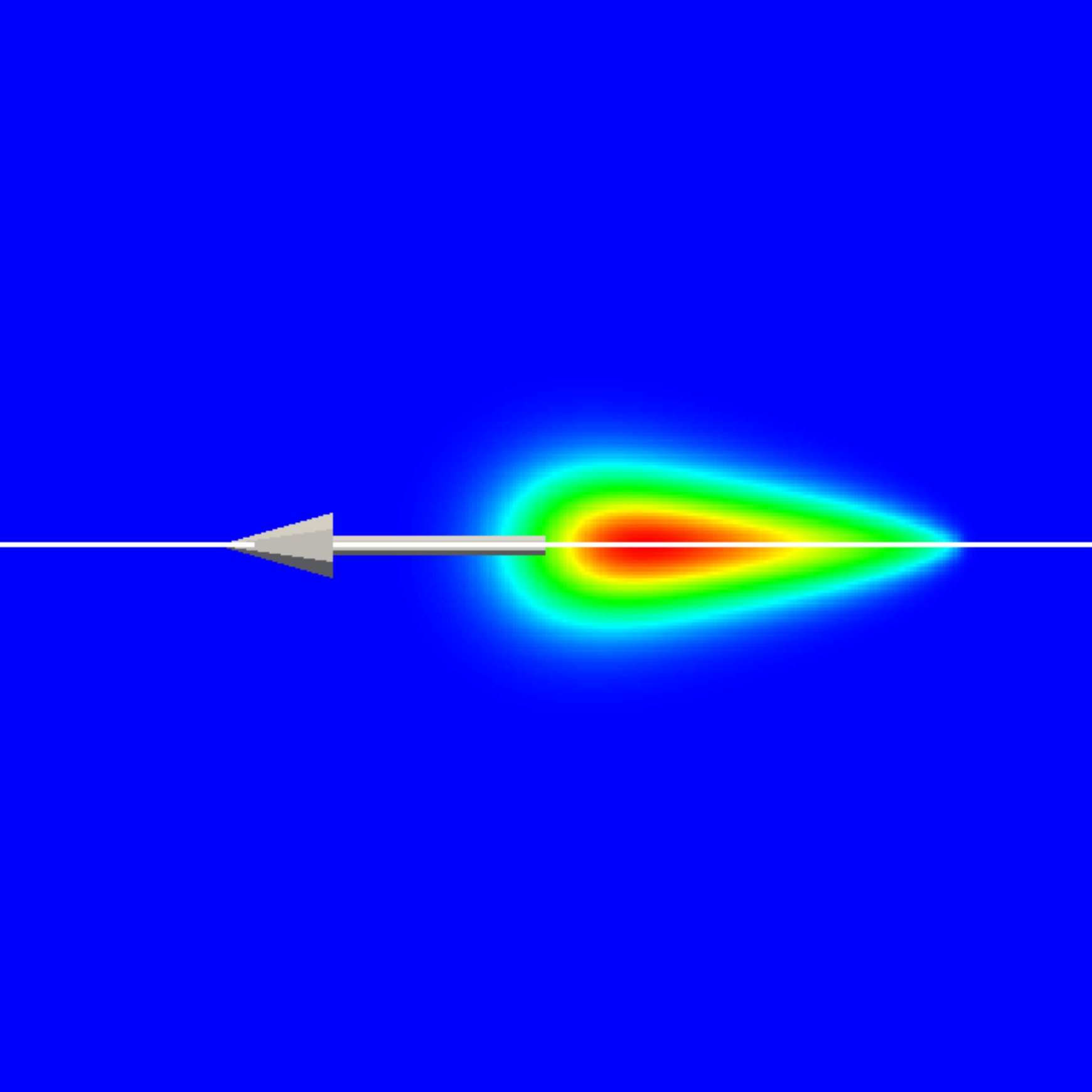} &\includegraphics[width=0.15\textwidth]{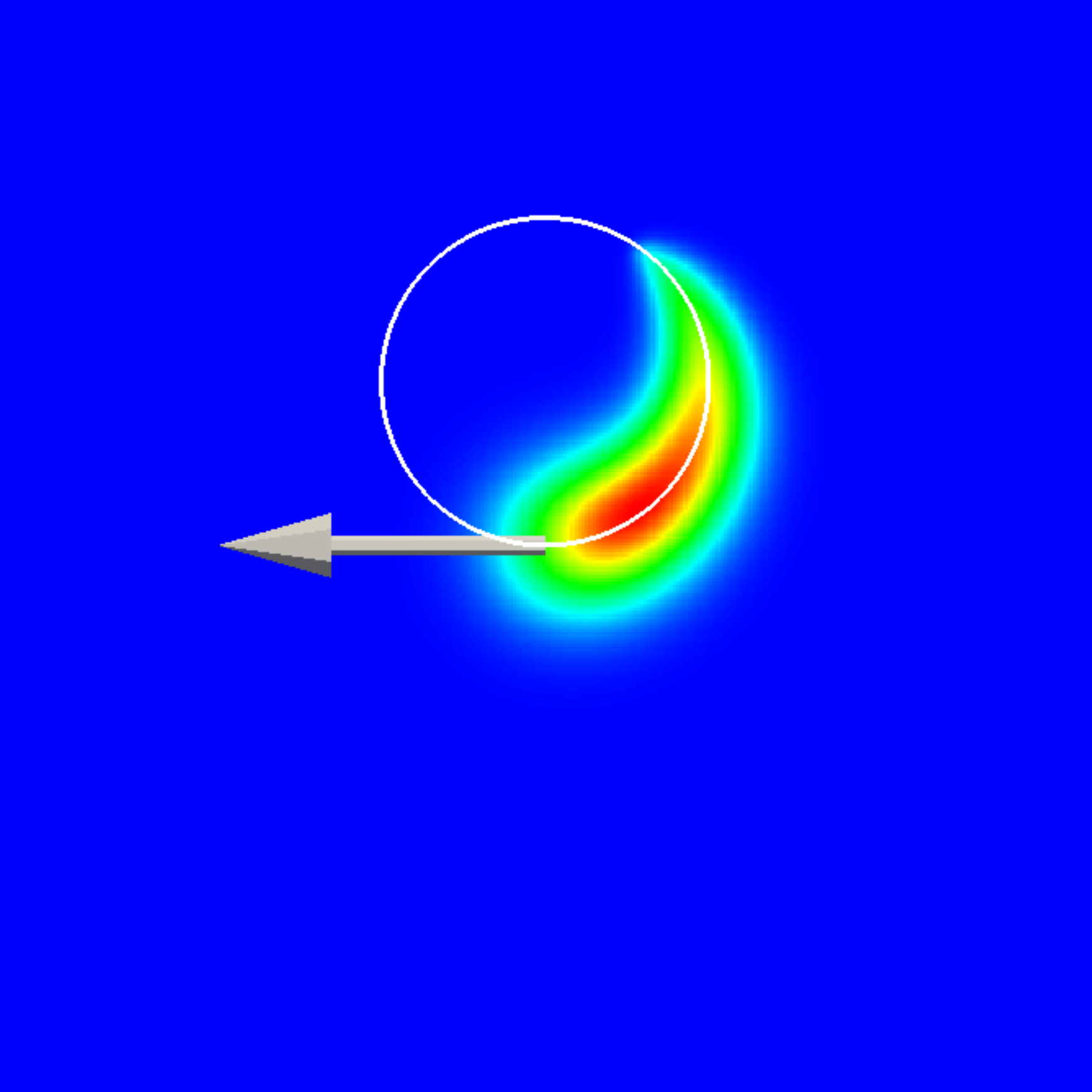}\\
\hline
concentration on a sphere in 3D & \includegraphics[width=0.15\textwidth]{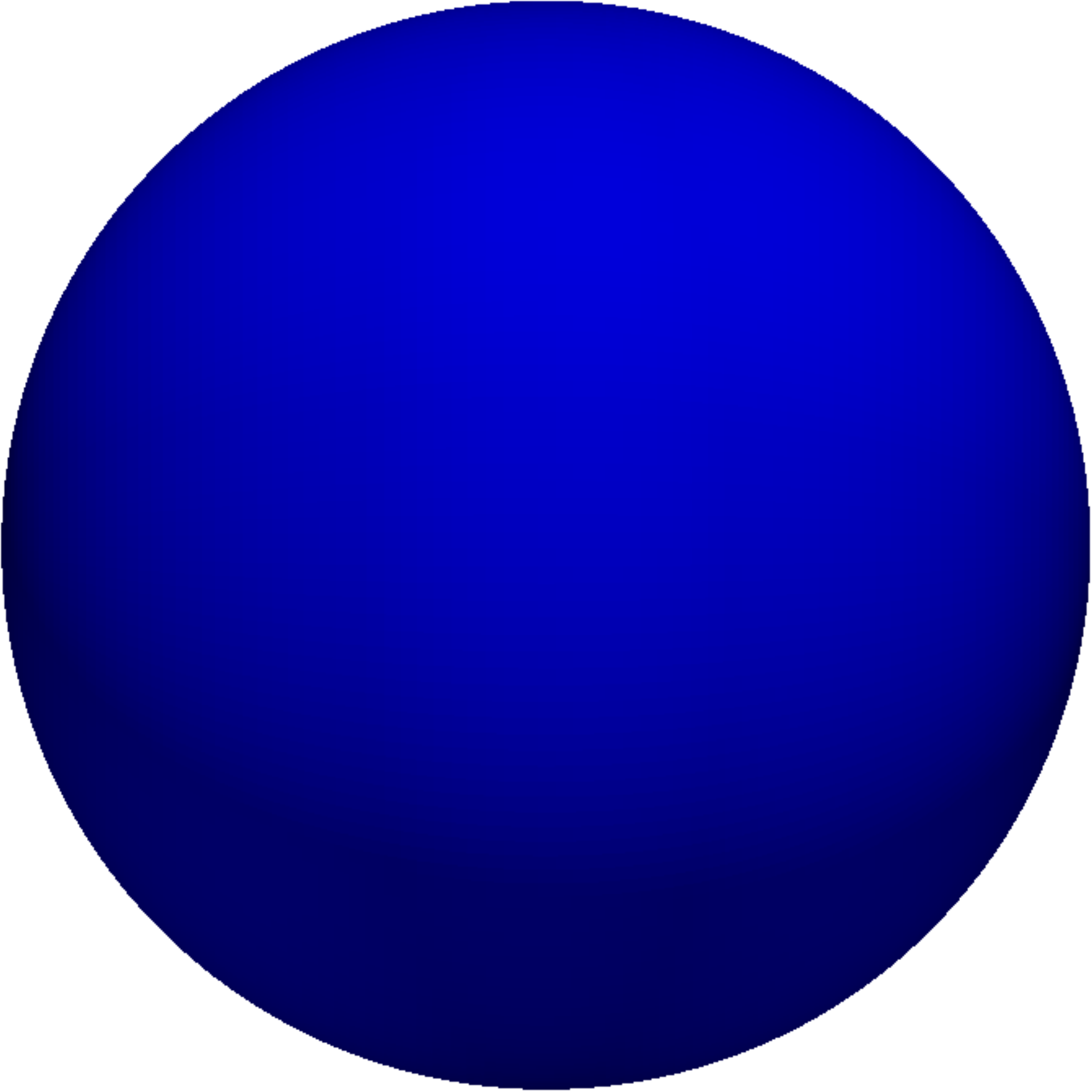} &\includegraphics[width=0.15\textwidth]{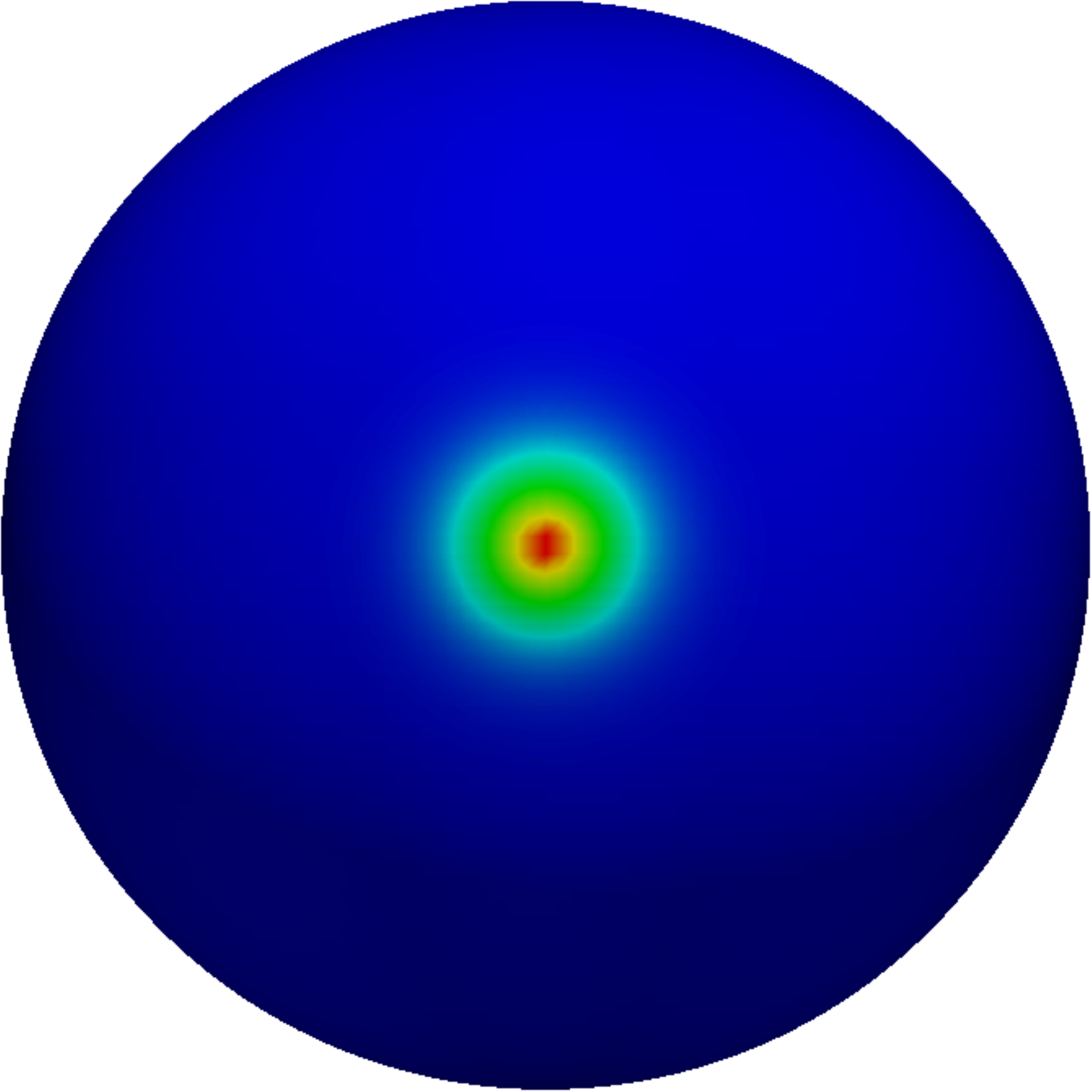} &\includegraphics[width=0.15\textwidth]{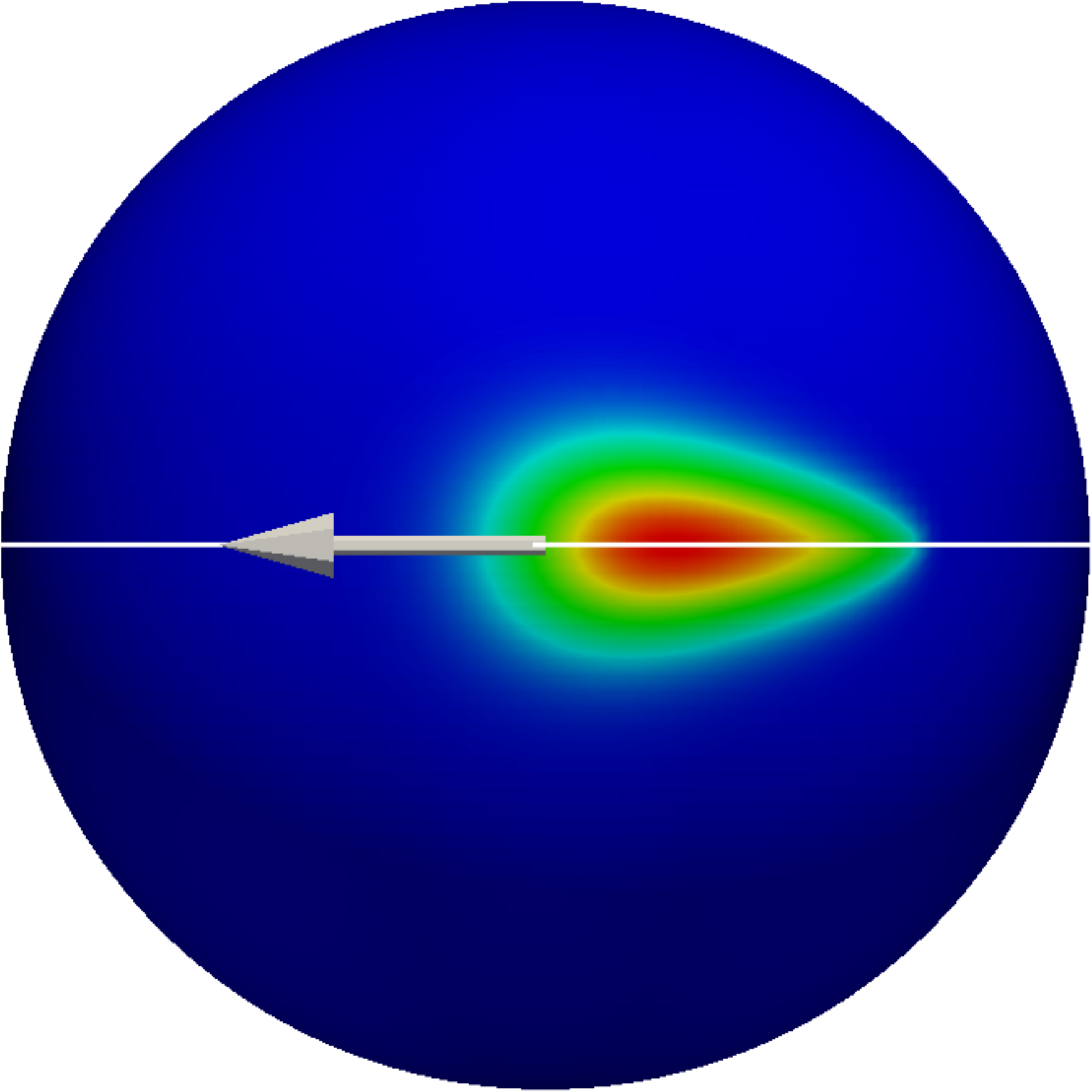} &\includegraphics[width=0.15\textwidth]{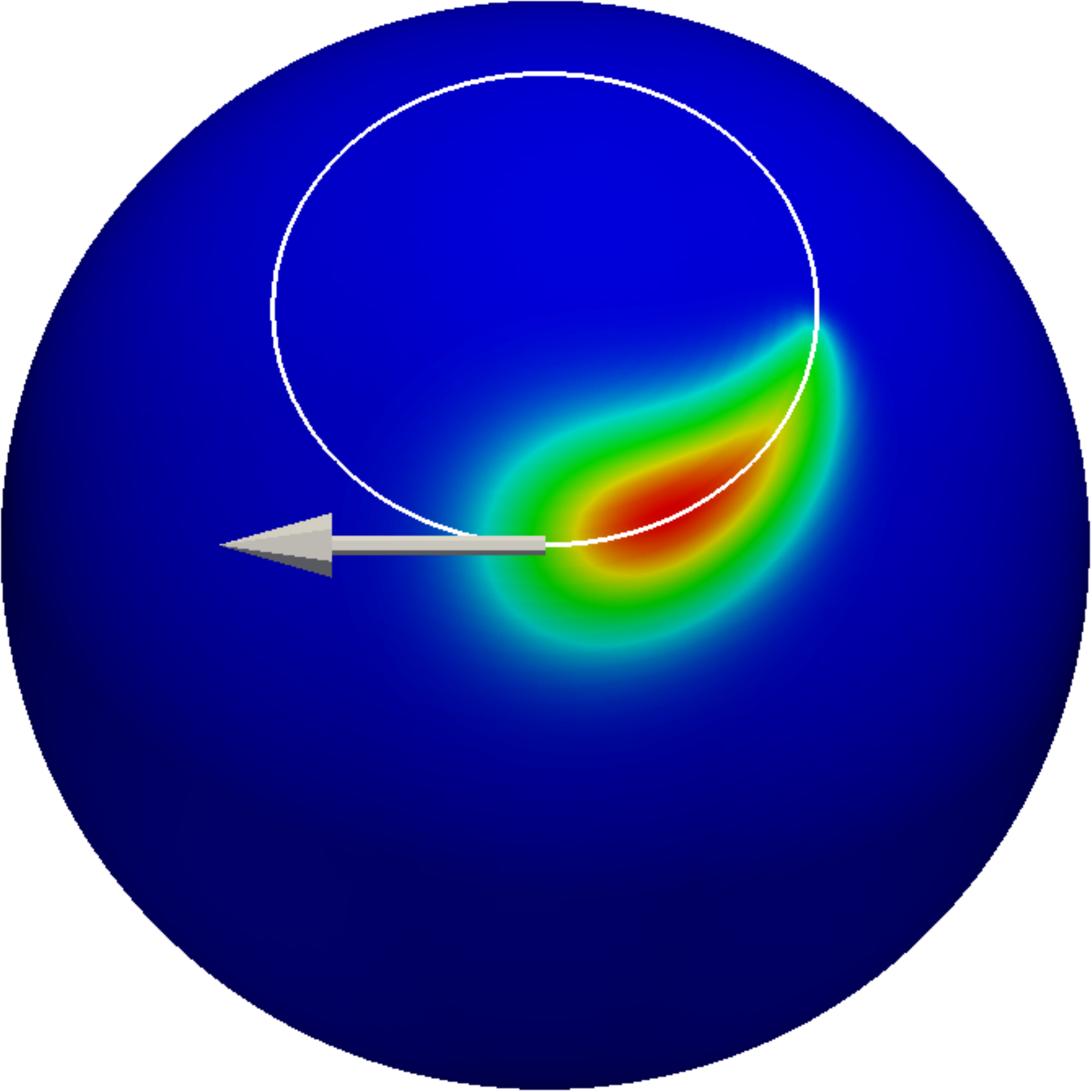}\\
\hline
trajectory in 2D & -- & stationary & straight & circle\\
\hline
locus of velocity orientations in 3D & -- & stationary & great circle & small circle \\
\hline
trajectory in 3D & stationary & straight & circle & helix
\end{tabular}
\caption{\label{concentrationtable} Analogy between the types of self-congruent solutions on a plane in 2D and the angular dependence of the types of self-congruent solutions in 3D.
Color code represents the concentration field.
White curves of the concentration field and arrows in columns C and D show the time evolution of the system:
The white curves show the trajectory of the concentration maximum (in comoving frame for the spherical case).
The white arrows show the instantaneous motion of the concentration maximum.
}
\end{figure*}

\subsection{3D case: spherical particle in a space}

Here we provide a brief description of the 3D case.
An explicit model showing the bifurcations presented here is analyzed in detail in the next Section.
The bifurcation sequence in 3D space follows the pattern described for 1D and 2D.
It is convenient to relate the angular dependence of the concentration field  in  3D  to  that in   2D, as shown in Fig. \ref{concentrationtable}.
For low enough $Pe$ the concentration field has the full rotational symmetry, which corresponds to a homogeneous concentration in 2D (Fig. \ref{concentrationtable}, column A).
The angular dependence of a given perturbation mode in 3D corresponds to a spherical harmonic of some degree.
All perturbations decay for low enough activity and the system is stationary in this case.
The spherical symmetry is broken at a critical activity $Pe_1$.
If the angular dependence of the mode that becomes unstable at $Pe=Pe_1$ corresponds to the first spherical harmonic, the bifurcation results in a motile self-congruent solution.
A circular spot of concentration excitation appears in the angular dependence, which corresponds to a stationary axisymmetric concentration field in 2D (Fig. \ref{concentrationtable}, column B).
The 3D particle moves along a fixed direction defined by the orientation of the spot (where the 2D analogue corresponds to a non motile solution).
The norm of the velocity scales as $(Pe-Pe_1)^{1/2}$ close to the transition point.
According to the 2D analysis, the concentration spot loses its axial symmetry and moves along a straight line at $Pe=Pe_2$.
The spherical counterpart of this effect is a non-circular spot of concentration moving along  the equator  (Fig. \ref{concentrationtable}, column C).
This corresponds to the case when the angular velocity of the corotational frame is orthogonal to the translational velocity, meaning that the particle moves along a circle.
The radius of the circle scales as $(Pe-Pe_2)^{-1/2}$ close to the transition point.
Finally, this solution becomes unstable with respect to the loss of the remaining mirror symmetry at $Pe=Pe_3$.
In 2D, the concentration spot moves along the particle periphery, corresponding to a  circular path for the particle.
In 3D this corresponds to a concentration spot moving along a small circle (i.e. a circle outside the equator; Fig. \ref{concentrationtable}, column D).
The 3D trajectory is a helix in this case.
The pitch of the helix scales as $(Pe-Pe_3)^{1/2}$ close to the transition point.

\section{\label{sect:analytical}Example: A simple exactly solvable model}
\subsection{Model formulation}
The above analysis shows how a helical trajectory can appear after a series of symmetry-breaking bifurcations for an isotropic motile particle.
The question however remains whether this series of bifurcations can be actually realized in practice.
To show that this is indeed the case, we constructed an exactly solvable model which manifests the desired trajectory types.
The model is taken as simple as possible and thus relies only on two harmonics of the concentration field, the first harmonic $c_i$ and the second one $c_{ij}$.
Here $c_i$ is a 3D vector and $c_{ij}$ is a 3D symmetric traceless tensor.
This Cartesian representation turns out to simplify notations in comparison to use of spherical coordinates.
The particle is taken as a unit sphere with a concentration field distributed on its surface.
The concentration field at point $\boldsymbol x$ on the sphere is given by
\begin{equation}
\label{concentration}
c(\boldsymbol{r})=c_ir_i+c_{ij}r_ir_j.
\end{equation}
We propose the following dynamics equations:
\begin{subequations}
\begin{align}
\label{dynamicsa}
	\dot c_i&=\sigma_1 c_i+\alpha_1 c_j^2c_i+\beta_1(c_k^2c_{ij}c_j-c_jc_kc_{jk}c_i)\\
\label{dynamicsb}
	\dot c_{ij}&=\sigma_2 c_{ij}+\beta_2(c_ic_j-\delta_{ij}c_k^2/3).
\end{align}
\end{subequations}
where $\delta_{ij}$ is the identity matrix.
We choose the propulsion velocity $\boldsymbol v$ of the particle as $v_i=c_i$.  As explained in the companion Letter \cite{misbahLetter} whenever the velocity is a linear  function of concentration  (and/or of  derivatives of $c$), the component of the swimming velocity, $v_i$, is linear in $c_i$ (first harmonic component).
The above equations are based on symmetry as they are invariant under rotations of the coordinate system.
The particular form of the last term of Eq. (\ref{dynamicsa}) is chosen in such a way that the equation for the amplitude of $c_i$ be independent of $c_{ij}$ to simplify the symbolic analysis.
This can easily be checked by multiplying the first equation by $c_i$.
The next Section presents a more generic model, which we analyze numerically with qualitatively the same results.

As we show below, the system (\ref{dynamicsa},\hyperref[dynamicsb]{b}) possesses the desired properties:

\begin{itemize}
\item Its solutions can be found analytically.
\item The stable solution goes from no motion to straight motion to circular motion to helical motion when $\sigma_1$ is increased and the other parameters are fixed. In other words, self-congruent solutions emerge naturally.
\end{itemize}

\subsection{Primary bifurcation}
The key to understanding the system (\ref{dynamicsa},\hyperref[dynamicsb]{b}) lies in the fact that the $\beta_1$ term in eq. (\ref{dynamicsa}) is orthogonal to $c_i$.
This leaves the evolution of the norm of $c_i$ independent of $c_{ij}$:
\begin{equation}
\label{norm}
	c_i\dot c_i=\sigma_1 c_i^2+\alpha_1 (c_i^2)^2.
\end{equation}
This is the classical form of a pitchfork bifurcation.
We choose $\alpha_1<0$ in order to ascertain  a supercritical bifurcation.
With this choice, we obtain that for $\sigma_1<0$ the stable solution is $c_i^2=0$, which corresponds to a non-motile case.
For $\sigma_1>0$, the stable solution is 
\begin{equation}
\label{c10}
	c_i^2=-\sigma_1/\alpha_1,
\end{equation}
which corresponds to a motile solution.
Since the norm dynamics of $c_i$ is intrinsic (independent of $c_{ij}$), we assume below that the norm of $c_i$ has already reached its stationary value defined by eq. (\ref{c10}).
The exact form of the motile solution depends only on  the orientational dynamics of $c_i$.

As we have seen before for a self-congruent solution the concentration field is stationary in a given frame (to be determined) which is related to the laboratory frame by a rigid translation and rotation. For this reason we find  it  more convenient to rewrite the equations in the  frame co-rotating  
 with $c_i$.
We call $\boldsymbol\omega$ the angular velocity (unknown for the moment) of the coordinate system which would keep the orientation of $c_i$ fixed.
In the corotating frame  the time derivative transforms as: 
\begin{equation}
\label{material}
\dot c(\boldsymbol r) \rightarrow \dot c(\boldsymbol r)-[\boldsymbol\omega\boldsymbol\times\boldsymbol{r}]\boldsymbol\cdot\boldsymbol\nabla^s c(\boldsymbol r),
\end{equation}
where $\boldsymbol\nabla^s$ is the surface gradient operator on the sphere.
Here $-[\boldsymbol\omega\boldsymbol\times\boldsymbol{r}]$ is the velocity field on the sphere measured in the corrotational frame.

Equation (\ref{material}) defines the change of equations (\ref{dynamicsa},\hyperref[dynamicsb]{b}) when going in the corrotational frame:
\begin{subequations}
\begin{align}
\label{dynamicsra}
	\dot c_i+\epsilon_{ijk}\omega_jc_k=\sigma_1 c_i+\alpha_1 c_j^2c_i+\beta_1(c_k^2c_{ij}c_j-c_jc_kc_{jk}c_i)\\
\label{dynamicsrb}
	\dot c_{ij}+\epsilon_{ikl}\omega_kc_{lj}+\epsilon_{jkl}\omega_kc_{li}=\sigma_2 c_{ij}+\beta_2(c_ic_j-\delta_{ij}c_k^2/3).
\end{align}
\end{subequations}

Setting the angular velocity to
\begin{equation}
\label{omega}
\omega_i=\beta_1\epsilon_{ijk}c_jc_{kl}c_l+\mu c_i
\end{equation}
cancels the $\beta_1$ term in eq. (\ref{dynamicsra}).
Here $\mu$ is a constant, the value of which does not affect $\dot c_i$ in eq. (\ref{dynamicsra}).
This happens because the vector $c_i$ remains invariant on any rotation about its direction.
We choose $\mu$ to cancel the rotation of $c_{ij}$ about $c_i$, as explained below.
This implies that $\mu$ is set to 0 except  when dealing with the helical case.

The choice (\ref{omega}) makes Eq. (\ref{dynamicsra}) trivial.
This reduces the analysis of the rotational dynamics of $c_i$ to the study of eq. (\ref{dynamicsrb}), where $\boldsymbol\omega$ and $|c_i|$ are defined by eqs. (\ref{c10}) and (\ref{omega}), respectively.
We choose the direction of $c_i$ as the $x$ axis in the co-rotational frame.

\subsection{\label{secondary}Secondary bifurcation}
We saw above that if only the first harmonic ($c_i$) is taken into account then there is a supercritical bifurcation from non-motile to a motile state (when $\sigma_1>0$; see Eq.(\ref{c10})).
This is referred to as the primary bifurcation.
The question we answer below is whether or not the system shows other types of trajectory (circular, helical), and if so under which conditions.
The circular solution results from the loss of stability of the straight solution, and we refer to this situation as the secondary bifurcation.

It is clear from (\ref{omega}) that the occurrence of a non-zero rotation of $c_i$ requires a coupling between first and second harmonics.
It is geometrically clear that if the direction of $c_i$ is taken as  the $x$ axis in the co-rotational frame, then a circular trajectory corresponds to $\boldsymbol\omega$ being orthogonal to $x$ axis.
For example, if $\boldsymbol\omega$ is along $z$, this means that $c_i$ moves along the equator in the $x,y$ plane in the lab frame.
The occurrence of non-zero $\boldsymbol\omega$ depends on symmetry of $c_{ij}$, as seen below.

The remaining procedure is the following: we first restrict the stationary solutions of $c_{ij}$ to the ones which respect a plane of symmetry, which corresponds to circular trajectories.
Then we analyze the linear stability of these solutions with respect to the loss of the mirror symmetry, which yields the transition point to the helical trajectories.
Finally, we relax the mirror symmetry in order to find the analytical expressions for the helical solutions.

Assuming the tensor $c_{ij}$ to be symmetric with respect to the $(x,y)$ plane, we can set $c_{xz}$ and $c_{yz}$ to 0.
Because $\omega$ is orthogonal to $c_i$, we set $\mu$ in eq. (\ref{omega}) to 0.
The only non-zero component of the angular velocity is $\omega_z$ in this case, given by
\begin{equation}
\label{omegaz}
\omega_z=\beta_{1} c^2_ic_{xy}
\end{equation}
Using (\ref{dynamicsrb}) and (\ref{omegaz})  yields the following equations for the diagonal terms:
\begin{subequations}
\begin{align}
\label{cxxdot}
	\dot c_{xx}&=\sigma_2 c_{xx}+2\beta_1 c^2_i c_{xy}^2 + \frac{2\beta_{2} c^2_i}{3},\\
\label{cyydot}
	\dot c_{yy}&=\sigma_2 c_{yy}-2\beta_1 c^2_i c_{xy}^{2} - \frac{\beta_{2} c^2_i}{3}.
\end{align}
\end{subequations}
The $\dot c_{zz}$ equation is linearly dependent on set (\ref{cxxdot},\hyperref[cyydot]{b}) by virtue of the zero-trace property of $c_{ij}$.
Choosing $\sigma_2$ to be negative is necessary to ensure linear  stability of $c_{ij}$ described by  Eqs. (\ref{cxxdot},\hyperref[cyydot]{b}).
Setting $\dot c_{xx}=\dot c_{yy}=0$ yields the expressions for $c_{xx}$ and $c_{yy}$ as functions of $c_{xy}$.
These expressions are substituted into the $\dot c_{xy}$ equation to provide
\begin{equation}
\label{cxydot}
\dot c_{xy}= \left(\frac{\beta_{1} \beta_{2} \left(c^{2}_{i}\right)^{2}}{\sigma_2} + \sigma_2\right)c_{xy}+\frac{4\beta_{1}^{2} \left(c^{2}_{i}\right)^2 }{\sigma_{2}}c_{xy}^{3}.
\end{equation}
Equation (\ref{cxydot}) corresponds to a pitchfork bifurcation, which is of supercritical type because the coefficient of the cubic term is negative for $\sigma_2<0$.
The coefficient of the linear term is negative for $c_i^2=0$ but can change its sign for large enough $c_i^2$, provided $\beta_1\beta_2<0$.
For $\beta_1\beta_2(c_i^2)^2<\sigma_2^2$, the stable solution is $c_{xy}=0$, which corresponds to an axisymmetric solution $c_{yy}=c_{zz}=-c_{xx}/2$.
The angular velocity is zero in this case and the particle moves along a straight line.
For $\beta_1\beta_2(c_i^2)^2>\sigma_2^2$, the stable solution is
\begin{equation}
\label{cxy0}
c_{xy}^2=- \frac{\beta_{2}}{4 \beta_{1}} - \frac{\sigma_{2}^{2}}{4 \beta_{1}^{2} \left(c^2_i\right)^2},
\end{equation}
which corresponds to non-zero $\boldsymbol\omega$.
This results in the concentration field rotating in the laboratory frame with angular velocity orthogonal to the propulsion velocity and defined by eqs. (\ref{omegaz}) and (\ref{cxy0}).
The trajectory is a circle in this case.
Indeed, the propulsion velocity $v_i=c_i$ is directed along $x$ and the rotation speed is along $z$ in the corotating frame.
This means that the propulsion velocity in the laboratory frame rotates along the equator contained in the plane $(x,y)$ with pulsation $\omega_z$.
It has thus two components $v_x \sim \sin (\omega_z t )$ and $v_y \sim \cos (\omega_z t)$, and the corresponding trajectory behaves as $x \sim \cos (\omega_z t )/\omega_z$ and $v_y \sim \sin (\omega_z t)/\omega_z$, which is a circle with radius proportional to $\omega_z ^{-1}$.
According to (\ref{omegaz})  $\omega_z$ is proportional to $c_{xy}$, which vanishes  as the square root of the distance from the bifurcation  point (defined by right hand side of  eq. (\ref{cxy0}) equal to zero).
This means that the circle radius diverges with square root singularity at the bifurcation point, as we find from our general consideration and explicitly in 2D in the companion Letter \cite{misbahLetter}.

\subsection{Tertiary bifurcation}
Now let us analyze the linear stability of the non-trivial solution of Eq. (\ref{cxydot}), which has led to a circular trajectory, with respect to the loss of the mirror symmetry. Recall that circular solution emerges at the secondary bifurcation. The loss of stability of the circular solution corresponds to a tertiary bifurcation. A helical solution emerges at this point as shown below. The circular solution corresponds to $c_{yz}=c_{xz}=0$ (mirror symmetry with respect to $(x,y)$ plane). The occurrence of a non-zero value of $c_{yz}$ or $c_{xz}$ breaks this symmetry and leads to a new trajectory. The
evolution equations for the $xz$ and $yz$ components of $c_{ij}$ in the corotating frame read (from Eq. (\ref{dynamicsrb}) with $\mu=0$)
\begin{subequations}
\begin{align}
\label{cxzdot}
	\dot c_{xz}&=\sigma_2c_{xz} + \beta_1c_i^2\left(c_{xy}c_{yz} + c_{xz} c_{yy} + 2 c_{xz} c_{zz}\right)\\
\label{cyzdot}
	\dot c_{yz}&=\sigma_2c_{yz} - 2\beta_1 c^2_i c_{xy} c_{xz}.
\end{align}
\end{subequations}
Substituting the steady-state values of $c_{yy}$, $c_{zz}$, and $c_{xy}$, we get that the determinant of the linear stability matrix is always zero.
This means that one of the eigenvalues of the linear stability matrix is zero.
This eigenvalue corresponds to the rotation of the problem about the $c_i$ direction.
Since one of the eigenvalues is zero, the trace of the linear stability matrix is equal to the second eigenvalue.
We therefore use the trace of (\ref{cxzdot},\hyperref[cyzdot]{b}) to analyze the stability of the circular motion:
\begin{equation}
\label{trace}
\frac{\partial\dot c_{xz}}{\partial c_{xz}}+\frac{\partial\dot c_{yz}}{\partial c_{yz}}=2\sigma_2+\beta_1c_i^2\left(c_{yy} + 2c_{zz}\right)=\frac{\beta_1 \beta_2 \left(c^2_i\right)^2}{2\sigma_2} + \frac{3 \sigma_2}{2},
\end{equation}
where the last equality is obtained by substituting the steady-state values of $c_{yy}$ and $c_{zz}$.
The circular motion thus becomes unstable for $(c_i^2)^2>3\sigma_2^2/(\beta_1\beta_2)$, which happens when $c_i^2$ (or, equivalently, $\sigma_1$) exceeds a critical value. 

It is possible to calculate the rotational velocity $\boldsymbol\omega$ analytically even without assumption of the $z\rightarrow -z$ symmetry of $c_{ij}$.
This is done by choosing the value of $\mu$ (setting $\omega_x$ in the corotating frame) in eq. (\ref{omega}) in such a way that $\dot c_{xz}=0$ in Eq. (\ref{dynamicsrb}).
Without loss of generality, we choose the $z$ axis such that $c_{xz}=0$.
The loss of $z\rightarrow -z$ symmetry is manifested as $c_{yz}\ne 0$ in this case.
The value of $\omega_x$ is given by
\begin{equation}
\label{omega1}
\omega_x=-\beta_{1}c^2_ic_{yz}.
\end{equation}

The evolution equation for $c_{yz}$ reads
\begin{equation}
\label{cyzdot2}
\dot c_{yz}=c_{yz} \left[\sigma_{2}-\beta_{1} c^2_i (c_{yy} - c_{zz})\right].
\end{equation}
Equation (\ref{cyzdot2}) shows two possibilities for fixed points: $c_{yz}=0$, which corresponds to the $z\rightarrow -z$ symmetric solution considered above in Section \ref{secondary} and a new possibility
\begin{equation}
\label{dc}
c_{yy} - c_{zz}=\frac{\sigma_2}{\beta_1 c^2_i}.
\end{equation}
Exploring this possibility yields
\begin{equation}
\label{cyz0}
c_{yz}^2=-\frac{\beta_{2}}{6\beta_{1}} - \frac{\sigma_{2}^{2}}{2 \beta_{1}^{2} \left(c^2_i\right)^{2}},
\end{equation}
where both signs of $c_{yz}$ are possible.
We also give the value $c_{xy}$ for the $c_{yz}\ne 0$ fixed point, which is necessary to evaluate the $\omega_z$ component of the angular velocity:
\begin{equation}
\label{cxy02}
c_{xy}^2=- \frac{\beta_{2}}{3 \beta_{1}} - \frac{\sigma_{2}^{2}}{2 \beta_{1}^{2} \left(c^2_i\right)^2}.
\end{equation}

The solution defined by eqs. (\ref{cyz0}) and (\ref{cxy02}) corresponds to helical motion according to the analysis in Section \ref{sect:SelfCongruent}.
Indeed, due to non-zero $c_{yz}$, $\omega^0_x\ne 0$, which implies that $\boldsymbol\omega^0$ and $\boldsymbol v^0$ be not orthogonal.
The pitch of the helix is defined by eq. (\ref{helix}).
The trajectories of the particle and the corresponding concentration fields can be obtained numerically, as shown in the next Section for a slightly different model.


\subsection{Bifurcation diagram}
The following assumptions were used to obtain all 3 possible bifurcations:
\begin{itemize}
\item $\sigma_2<0$
\item $\beta_1\beta_2<0$
\item $\alpha_1<0$
\end{itemize}
These assumptions guarantee that all bifurcations are of supercritical pitchfork type.
The final expressions for the translational and rotational velocities of the comoving and corotating frame read
\begin{equation}
\label{v0}
v^0=\begin{cases}
	0&\textrm{ for }\sigma_1\le 0\\
	\sqrt{-\frac{\sigma_1}{\alpha_1}}&\textrm{ for } \sigma_1>0,
\end{cases}
\end{equation}
\begin{equation}
\label{omegaperpendicular}
\omega_\bot^0=\begin{cases}
	0&\textrm{ for }\sigma_1\le \frac{\alpha_1\sigma_2}{\sqrt{-\beta_1\beta_2}}\\
	\sqrt{-\frac{\sigma_2^2}{4} - \frac{\beta_1\beta_2\sigma_1^2}{4\alpha_1^2}}&\textrm{ for }
		\frac{\alpha_1\sigma_2}{\sqrt{-\beta_1\beta_2}}<\sigma_1<\frac{\alpha_1\sigma_2\sqrt{3}}{\sqrt{-\beta_1\beta_2}}\\
	\sqrt{-\frac{\sigma_2^2}{2} - \frac{\beta_1\beta_2\sigma_1^2}{3\alpha_1^2}}&\textrm{ for }
		\sigma_1>\frac{\alpha_1\sigma_2\sqrt{3}}{\sqrt{-\beta_1\beta_2}},
\end{cases}
\end{equation}
\begin{equation}
\label{omegaparallel}
\omega_\parallel^0=\begin{cases}
	0&\textrm{ for }\sigma_1\le\frac{\alpha_1\sigma_2\sqrt{3}}{\sqrt{-\beta_1\beta_2}}\\
	\sqrt{-\frac{\sigma_2^2}{2}-\frac{\beta_1\beta_2\sigma_1^2}{6\alpha_1^2}}&\textrm{ for }
		\sigma_1>\frac{\alpha_1\sigma_2\sqrt{3}}{\sqrt{-\beta_1\beta_2}},
\end{cases}
\end{equation}
where $\boldsymbol\omega_\bot^0$ and $\boldsymbol\omega_\parallel^0$ are the components of the $\boldsymbol\omega^0$ perpendicular and parallel to $\boldsymbol v^0$, respectively.
As shown in Section \ref{sect:SelfCongruent}, the case $v^0=0$, $\omega^0=0$ corresponds to a stationary particle, the case $v^0>0$, $\omega^0=0$ corresponds to straight motion, the case $v^0>0$, $\omega^0_\bot>0$, $\omega^0_\parallel=0$ corresponds to a circular motion and the case $v^0>0$, $\omega^0_\bot>0$, $\omega^0_\parallel>0$ corresponds to a helical motion.

\begin{figure}
\begin{center}
\includegraphics[width=0.9\columnwidth]{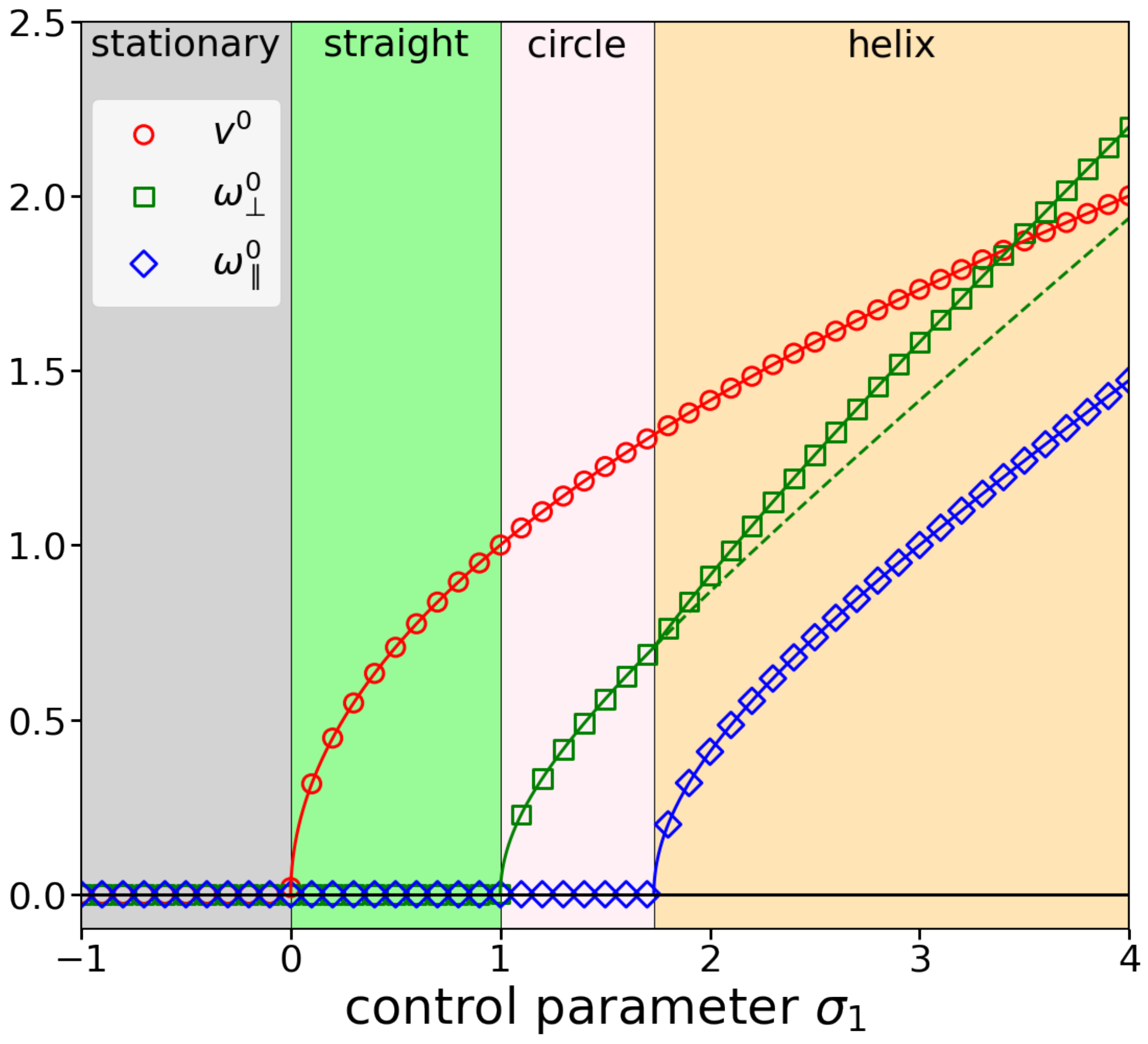}
\caption{\label{velocity}Saturation values of velocity $v^0$ and angular velocity $\omega^0$ of the comoving and corotating reference frame for a self-propelling particle described by eqs. (\ref{dynamicsa},\hyperref[dynamicsb]{b}).
The components of angular velocity along the velocity and orthogonal to it are shown as $\omega_\parallel^0$ and $\omega_\bot^0$, respectively.
$\sigma_2=-1$, $\alpha_1=-1$, $\beta_1=1$, $\beta_2=-1$.
Symbols are full numerical solution of equations (\ref{dynamicsa},\hyperref[dynamicsb]{b}).
Solid lines are analytical expressions.
The dashed line is the continuation of the circular solution in the helical phase.
The color regions denote the trajectory type.
}
\end{center}
\end{figure}

Figure \ref{velocity} shows expressions (\ref{v0}), (\ref{omegaperpendicular}) and (\ref{omegaparallel}) plotted as functions of $\sigma_1$.
We use direct numerical simulations of Eqs. (\ref{dynamicsa},\hyperref[dynamicsb]{b}) to validate the analytical results.
The details of the numerical procedure are given in the next section.
As can be seen, the analytical results match the numerical simulations exactly.
This confirms that the full dynamics of the system (\ref{dynamicsa},\hyperref[dynamicsb]{b}) relaxes to the appropriate self-congruent solution for the explored parameters.

We summarize the results in a table (Table \ref{motion}), showing the symmetry of the concentration field, the trajectory, and the concentration dynamics for different $\sigma_1$.

\begin{table*}
\begin{center}
\begin{tabular}{c|c|c|c}
$\sigma_1$ range & concentration symmetry & concentration maximum trajectory & trajectory \\
\hline
$\sigma_1<0$ & $O(3)$ (spherical) & -- & stationary \\
$0<\sigma_1<\frac{\alpha_1\sigma_2}{\sqrt{-\beta_1\beta_2}}$ & $C_{\infty v}$ (axial+mirror) & stationary & straight \\
$\frac{\alpha_1\sigma_2}{\sqrt{-\beta_1\beta_2}}<\sigma_1<\frac{\alpha_1\sigma_2\sqrt{3}}{\sqrt{-\beta_1\beta_2}}$ & $C_s$ (mirror) & great circle (equator) & circle \\
$\frac{\alpha_1\sigma_2\sqrt{3}}{\sqrt{-\beta_1\beta_2}}<\sigma_1$ & $C_1$ (trivial) & small circle & helix
\end{tabular}
\caption{\label{motion}Types of particle dynamics for different values of $\sigma_1$.}
\end{center}
\end{table*}

\section{Direct numerical solution}
Section \ref{sect:analytical} deals with a system of equations (system (\ref{dynamicsa},\hyperref[dynamicsb]{b})) which has a special property: the evolution of $c_i^2$ is completely decoupled from the rest of the dynamics.
We show here that $c_i^2$ relaxes to a time-independent value even if it is coupled to $c_{ij}$ and the orientation of $c_i$.
To highlight the genericity of the presented results, we write below (as we did in 2D in the companion Letter \cite{misbahLetter}) the evolution equations for the first and second harmonics based on symmetry only (invariance under 3D rotations). To the leading order we have
\begin{subequations}
\begin{align}
\label{numdynamicsa}
	\dot c_i&=\sigma_1 c_i+\alpha_1 c_j^2c_i+\beta_1c_{ij}c_j\\
\label{numdynamicsb}
	\dot c_{ij}&=\sigma_2 c_{ij}+\beta_2(c_ic_j-\delta_{ij}c_k^2/3).
\end{align}
\end{subequations}
Solving the differential equations (\ref{numdynamicsa},\hyperref[numdynamicsb]{b}) for given values of parameters yields the functions $c_i(t)$ and $c_{ij}(t)$.
The following procedure is used to extract the values $\omega^0_\bot$, and $\omega^0_\parallel$.
We define 3 mutually orthogonal unit vectors $\boldsymbol e^1$, $\boldsymbol e^2$, and $\boldsymbol e^3$, corotating with the concentration field:
\begin{subequations}
\begin{align}
\label{e1}
e^1_i&=\frac{c_i}{|c_i|},\\
\label{e2}
\boldsymbol e^2&=\frac{\boldsymbol e^1\boldsymbol\times\boldsymbol b}{|\boldsymbol e^1\boldsymbol\times\boldsymbol b|},\textrm{ where } b_i=c_{ij}c_j,\\
\label{e3}
\boldsymbol e^3&=\boldsymbol e^1\boldsymbol\times\boldsymbol e^2.
\end{align}
\end{subequations}
There are situations in which the vector $\boldsymbol e^2$ is ill-defined, meaning that the vector $b_i=c_{ij}c_j$ is parallel to $c_i$.
This only happens when the trajectory is not helical and we can set effectively $\omega_\parallel$ to 0 in this case.
The time derivatives of vectors $\boldsymbol e^i$ define the rotation of the concentration field.
Indeed, it is evident that a simultaneous rotation of $c_i$ and $c_{ij}$  leads to the same rotation of $\boldsymbol e^i$.
The vectors $\boldsymbol e^k$ with $k\in\{1,2,3\}$ satisfy the evolution equations $\dot{\boldsymbol e}^k=\boldsymbol\omega\boldsymbol \times \boldsymbol e^k$ for purely rotational dynamics.
This implies that
\begin{equation}
\label{omega2}
\sum\limits_{k=1}^3 (\dot {\boldsymbol e}^k)^2=\sum\limits_{k=1}^3 \left[\omega^2(e^k)^2-(\boldsymbol e^k\boldsymbol \cdot\boldsymbol\omega)^2\right]=2\omega^2,
\end{equation}
where we have used that the vectors $\boldsymbol e^k$ form an orthonormal basis.
Hence we compute the angular velocities as
\begin{equation}
\label{omeganum}
\omega_\bot=|\dot{\boldsymbol e}^1|,\,\,\,\omega_\parallel=\sqrt{\frac{(\dot {\boldsymbol e}^2)^2+(\dot {\boldsymbol e}^3)^2-(\dot {\boldsymbol e}^1)^2}{2}},
\end{equation}
using that ${\boldsymbol e}^1$ is parallel to $c_i$.

\begin{figure}
\begin{center}
\includegraphics[width=0.9\columnwidth]{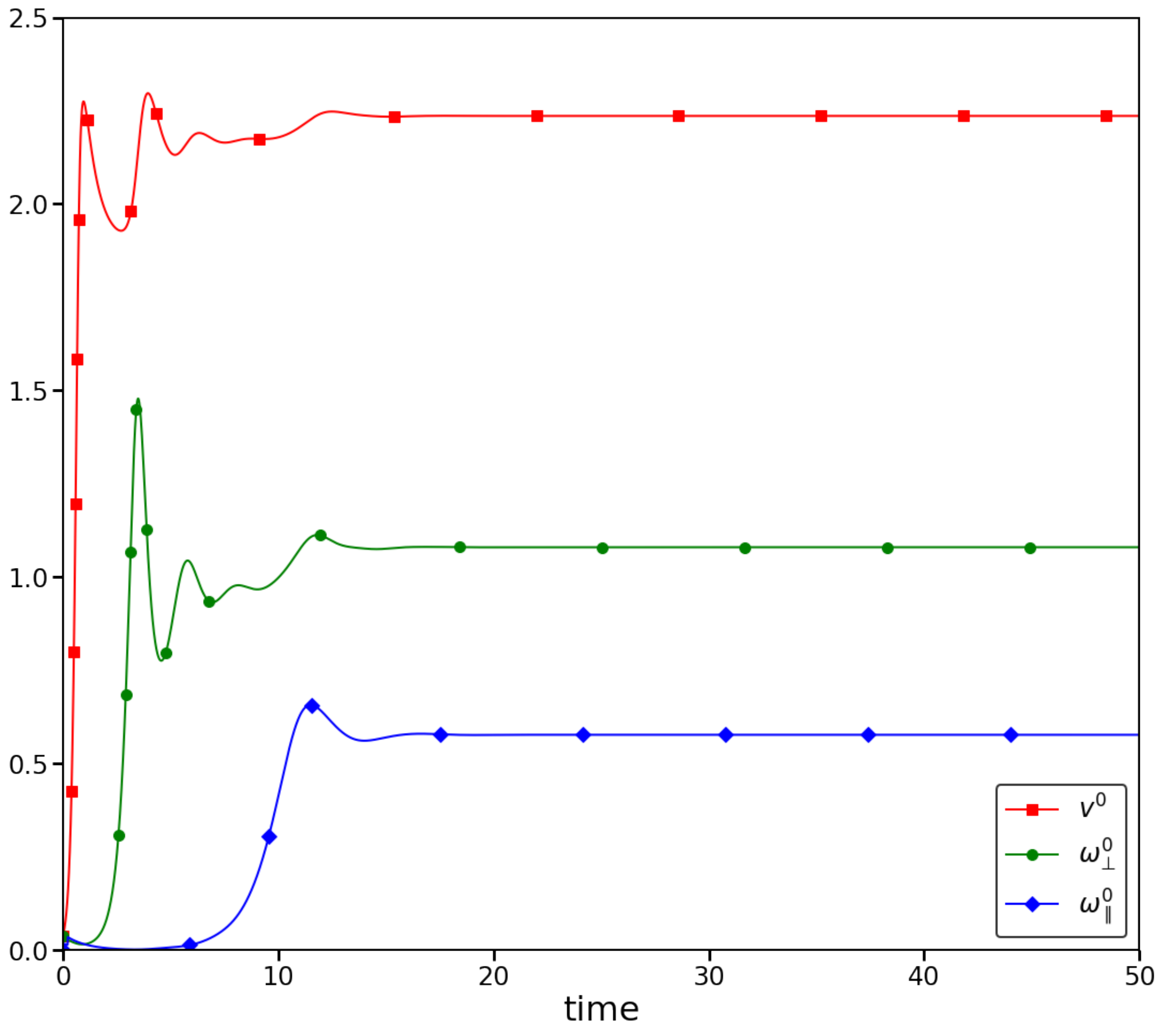}
\caption{\label{saturation}Velocity $v^0$ and angular velocity $\omega^0$ of the comoving and corotating reference frame for a self-propelling particle described by eqs. (\ref{numdynamicsa},\hyperref[numdynamicsb]{b}) as a function of time.
The components of angular velocity along the velocity and orthogonal to it are shown as $\omega_\parallel^0$ and $\omega_\bot^0$, respectively.
$\sigma_1=6$, $\sigma_2=-1$, $\alpha_1=-1$, $\beta_1=1$, $\beta_2=-1$.
Solid lines are full numerical solution of equations (\ref{numdynamicsa},\hyperref[numdynamicsb]{b}).
Symbols have no special meaning other than helping to distinguish the curves.
}
\end{center}
\end{figure}

First we show that the velocity and angular velocity of the comoving and corotating frame tend to a fixed value after the initial transients have decayed.
The results are presented in Fig. \ref{saturation} for a given set of parameters corresponding to the helical phase.
As can be seen, the system tends to a self-congruent dynamics after an initial transient. $v_0$ is given by the amplitude of $\bf c$ (first harmonic), and we see that after transients the amplitude tends to a constant, meaning that the choice of the exactly solvable model (\ref{dynamicsa},\hyperref[dynamicsb]{b}) (where we assumed $c_i^2$ is conserved) was legitimate.

\begin{figure}
\begin{center}
\includegraphics[width=0.9\columnwidth]{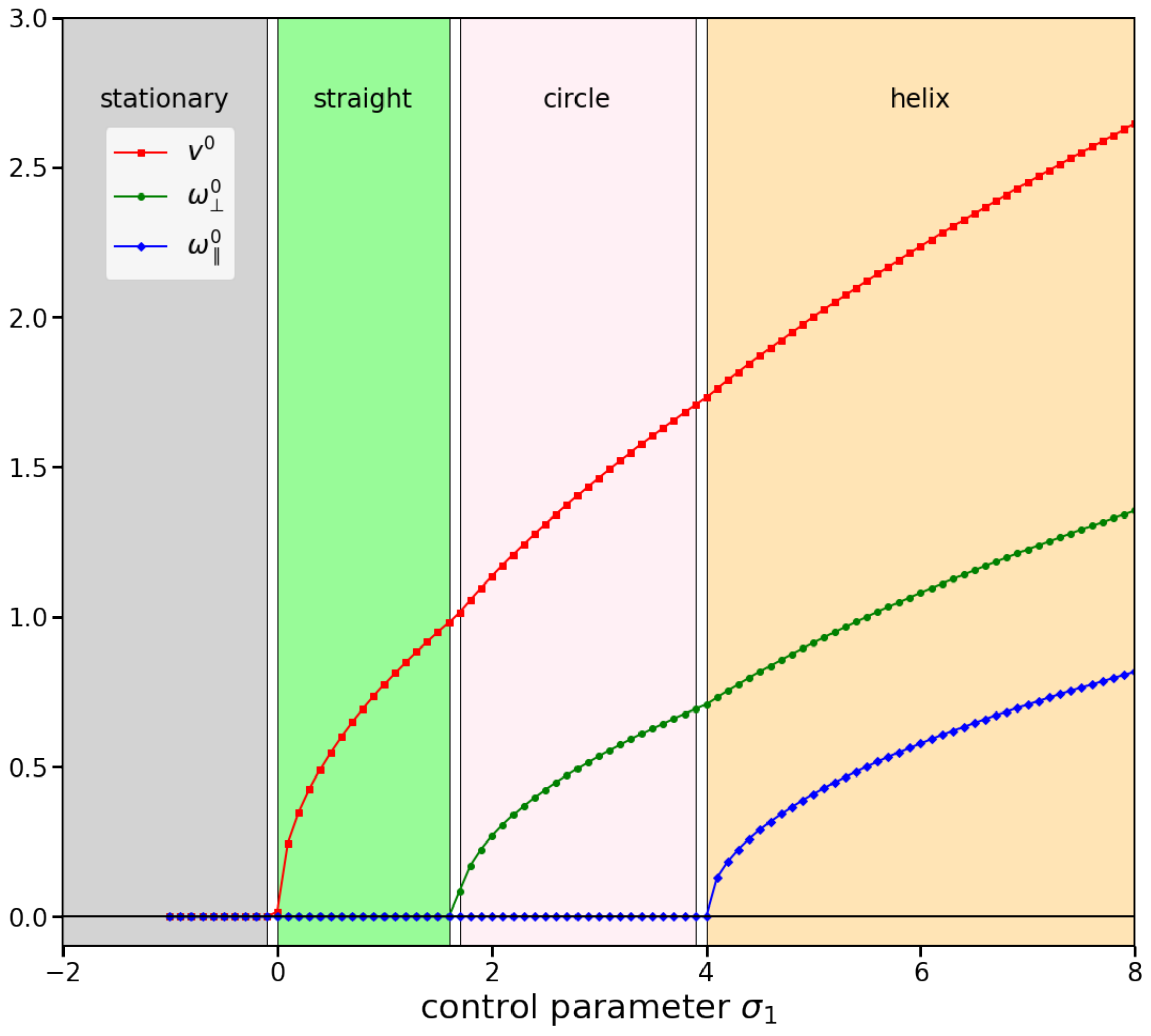}
\caption{\label{velocity2}Velocity $v^0$ and angular velocity $\omega^0$ of the comoving and corotating reference frame for a self-propelling particle described by eqs. (\ref{numdynamicsa},\hyperref[numdynamicsb]{b}).
The components of angular velocity along the velocity and orthogonal to it are shown as $\omega_\parallel^0$ and $\omega_\bot^0$, respectively.
$\sigma_2=-1$, $\alpha_1=-1$, $\beta_1=1$, $\beta_2=-1$.
Solid lines are full numerical solution of equations (\ref{numdynamicsa},\hyperref[numdynamicsb]{b}).
The color regions denote the trajectory type.
Symbols have no special meaning other than helping to distinguish the curves.
}
\end{center}
\end{figure}

Next, we plot the saturation values of $v^0$, $\omega_\parallel^0$, and $\omega_\bot^0$ as functions of $\sigma_1$.
The results are shown in Fig. \ref{velocity2}.
As can be seen, the bifurcation diagram is similar to the one discussed in the previous Section (Fig. \ref{velocity}).

\begin{figure*}
\begin{center}
\includegraphics[width=0.25\columnwidth]{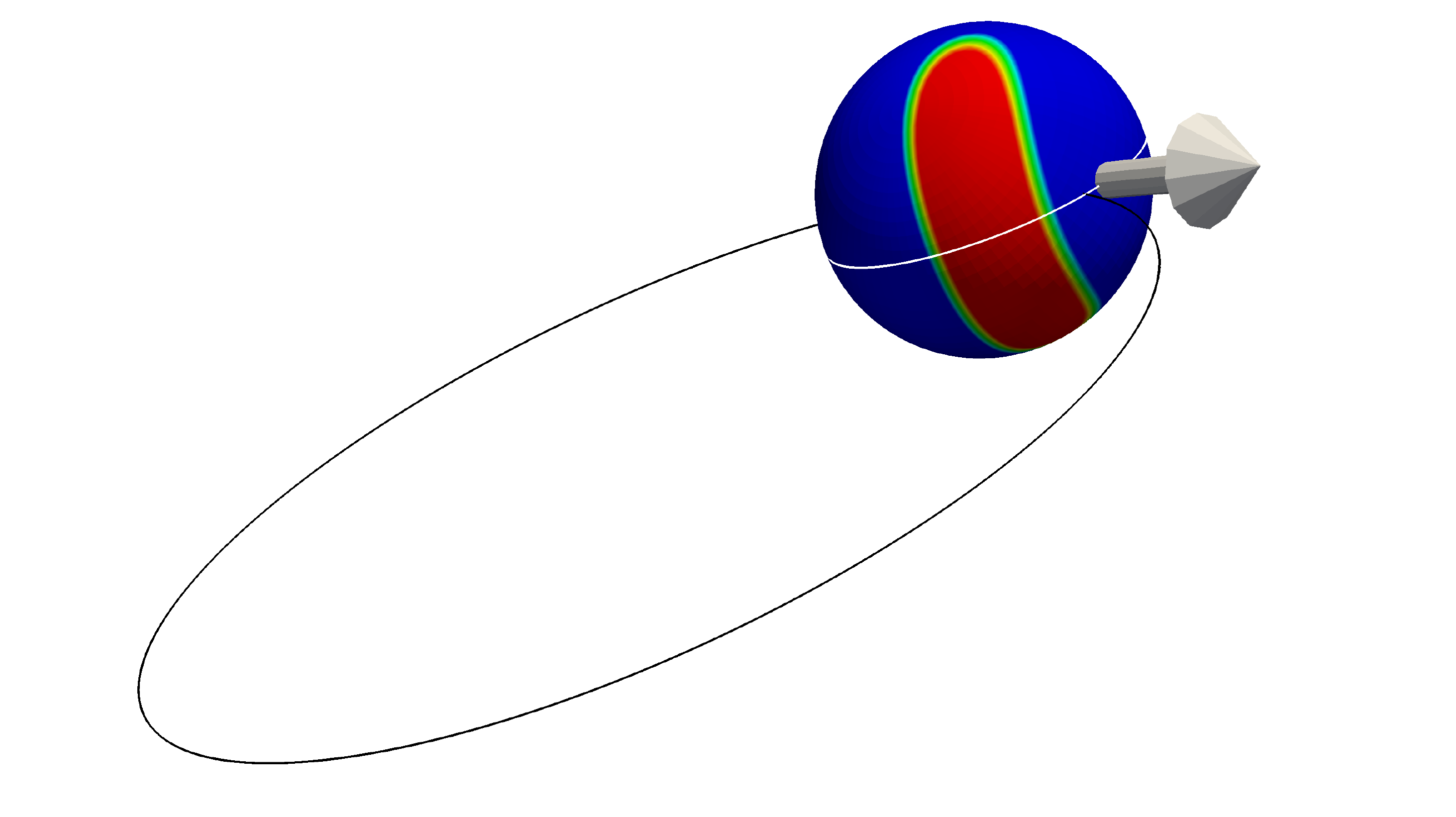}
\includegraphics[width=0.4\columnwidth]{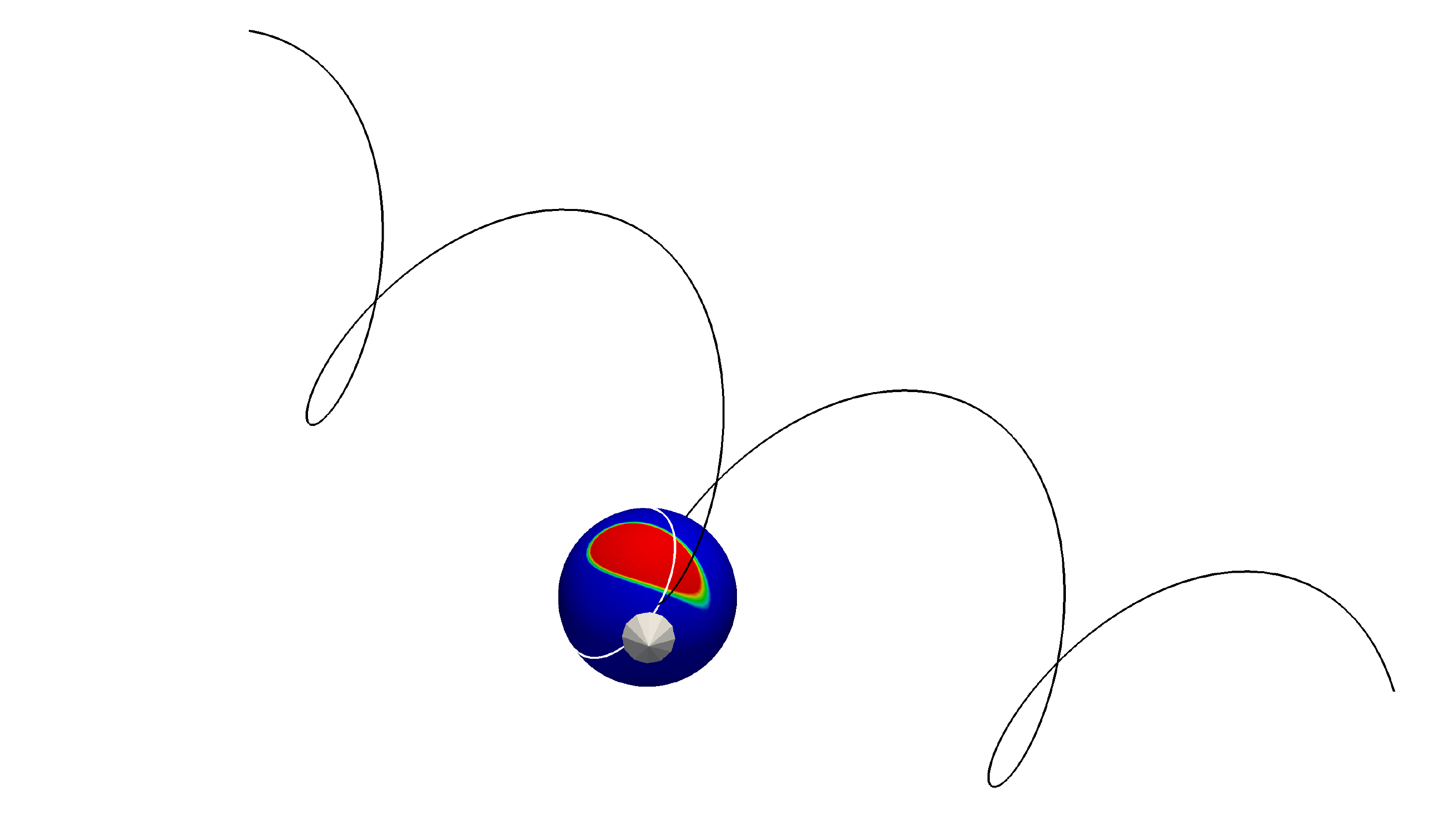}
\includegraphics[width=0.3\columnwidth]{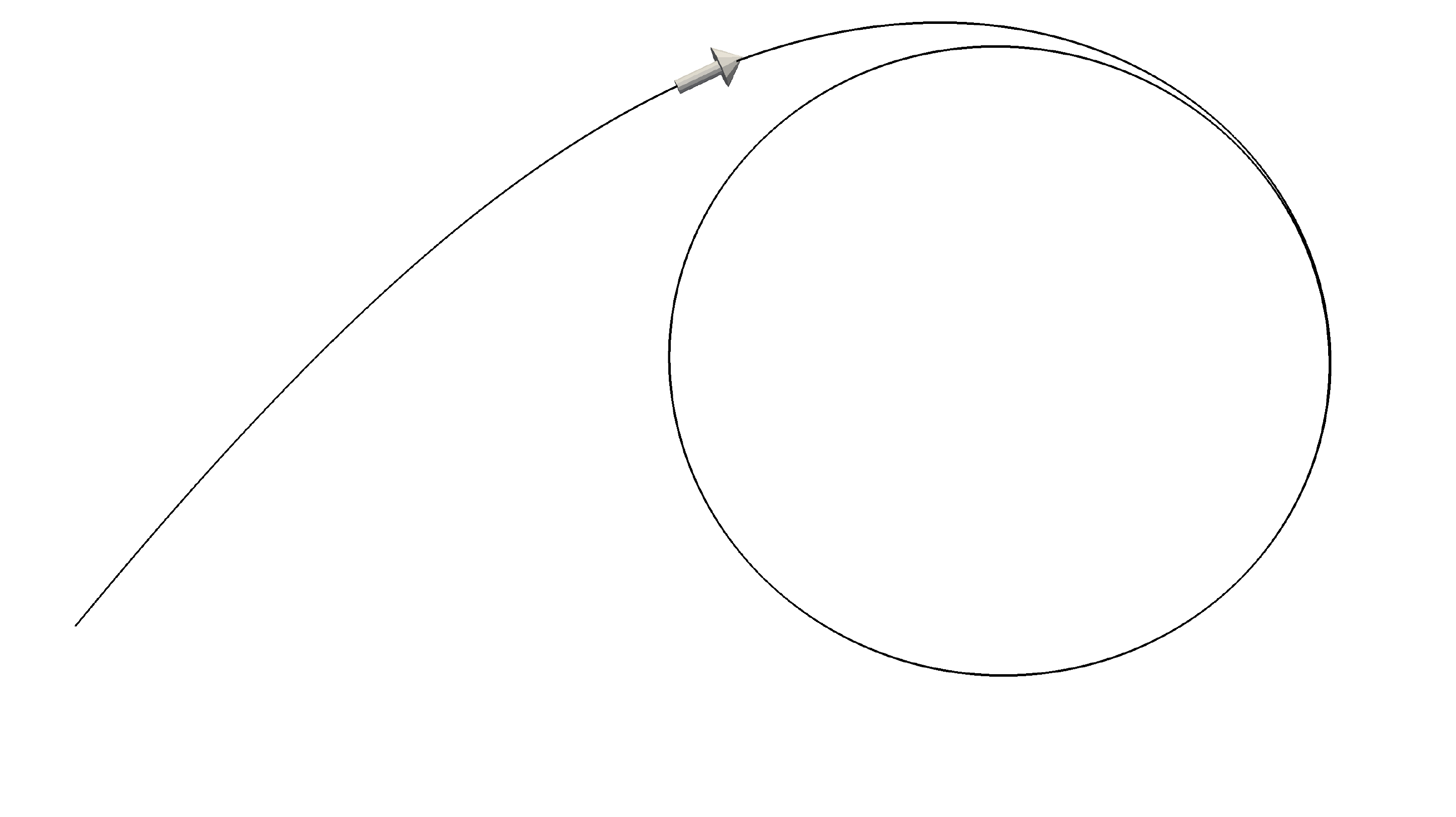}
\caption{\label{trajectories} Characteristic concentration distributions and particle trajectories.
Time-dependent solutions of eqs. (\ref{numdynamicsa},\hyperref[numdynamicsb]{b}).
$\sigma_2=-1$, $\alpha_1=-1$, $\beta_1=1$, $\beta_2=-1$.
Left: circular trajectory $\sigma_1=3$, Center: helical trajectory $\sigma_1=5$, Right: transient trajectory from straight to circular motion $\sigma_1=2$.
Color code: concentration field, black curve: particle trajectory, white curve on particle surface: locus of velocity orientations, arrow: instantaneous velocity.
The color code is scaled in a way that highlights the spot of high concentration.
}
\end{center}
\end{figure*}

We use the numerical solutions $c_i(t)$ and $c_{ij}(t)$ to plot the dynamics of the concentration field (defined by eq.(\ref{concentration})) and the trajectories of the particle in 3D, as shown in Fig.\ref{trajectories}.
The non-motile solution is characterized by zero concentration field (not shown in Fig.\ref{trajectories}).
The concentration field for straight motion has an axisymmetric spot of high concentration, the orientation of which remains constant with respect to the particle center.
Left panel of Fig. \ref{trajectories} shows the concentration field for the case of circular trajectory.
As can be seen, the concentration field is symmetric with respect to a plane passing through the center of the particle.
The rotation of the concentration field is such that the center of the high-concentration spot (which corresponds to the orientation of the particle velocity) moves along the equator (white curve on the particle surface).
For the helical case (central panel of Fig. \ref{trajectories}), the spot of high concentration is not symmetric and rotates in such a way that the velocity direction follows a closed loop corresponding to a small circle on the particle surface.
Finally we also present the transient behavior of the system, when the particle initially starts to move along a straight line until the symmetry breaking instability develops and the trajectory becomes circular (right panel of Fig. \ref{trajectories}).
This corresponds to growth of $\boldsymbol \omega^0$ from 0 to its saturation value, while $\boldsymbol v^0$ remains close to a constant and orthogonal to $\boldsymbol \omega^0$.
We can see that the trajectory spirals in to the circle in this case.


\section{Discussion}

The presented analysis shows that straight, circular, and helical trajectories in systems powered by a concentration field (like 
phoretic systems, and motile systems assisted by acto-myosin kinetics)
belong to a general class of self-congruent solutions and can emerge through a series of pitchfork bifurcations from a stationary solution.
Our model is very general and relies on the motility of the particle being related to a concentration field which can be distributed on the particle surface and/or in the media inside or outside it.
The shape of the particle is taken rotationally invariant and the breaking of the rotational symmetry of the system occurs spontaneously in the concentration field as the particle activity is increased.
An important feature of our model is that the position of the particle and its propulsion velocity are incorporated implicitly, similarly to the phase-field models.
That is, the model applies as long as the shape or position of the particle are not independent but are a function of the concentration field.
This is true, for example, even for deformable particles, as long as the shape relaxation is fast enough for adiabatic elimination to make sense.
Furthermore, the particle does not need to be a physical entity and the model pertains as well to soliton-like solutions when concentration peaks move in a homogeneous medium.

This generality of the model makes it applicable for a very diverse set of physical situations:
The prototypical example is an autophoretic droplet\cite{MLB13,morozov2019nonlinear,schmitt2013swimming,Hu2019}.
Another example is the model of cell motility, where the activity is driven by myosin motors distributed in the cortex along the surface of a spherical cell and/or  inside it \cite{hawkins2011spontaneous,Voituriez2016,Farutin2019}. 
Our model is still applicable if the cell is not spherical, as long as the shape of the cell can be reconstructed from the concentration field and is a sphere for low enough motility.
Another example is the dynamics of myosin concentration in tissues, where our model can be used to describe the motion of localized concentration peaks \cite{Negro2019,BLANCHARD201878}.

This study considers the case when the symmetry breaking occurs by a steady pitchfork bifurcation.
We considered the case of supercritical bifurcation when the solution changes continuously at the critical point.
The transition is discontinuous for subcritical bifurcations.
In this case, it is generally not possible to predict the properties of the motile solutions by analyzing the dynamics close to the non-motile branch. However, for a weakly subscritical bifurcation, a systematic analysis similar to the present one can be adopted.

Another possibility is Hopf bifurcation.
The solutions show periodic oscillations in this case.
For example, a 1D phoretic system undergoing Hopf bifurcation would show a back-and-forth motion of the particle.
The oscillations can arise due to a strong coupling of two different spatial modes.
Another possibility is coupling between concentration dynamics of two chemical species, such as actin and myosin\cite{Farutin2019}.
In 2D, Hopf bifurcation can happen either as the primary (translational modes) or the secondary (angular) instability.
There are two translational degrees of freedom in 2D, which are both excited in the case of the primary bifurcation.
The non-linear terms set the interaction between the oscillations along two orthogonal directions.
There are two possible solutions in this case, which correspond to planar and circular polarization.
The first case is similar to the 1D situation with back-and-forth motion of the particle, while the second case shows circular trajectories.
The radius of the circle grows continuously from zero at the instability point in this case.
Hopf bifurcation as the secondary instability leads to periodic oscillations of the velocity orientation, which leads to meandering \cite{Hu2019}. It is hoped to investigate this matter in the future.

Here we have considered that the spectrum is discrete, meaning that the system size is finite (albeit arbitrary large).  For a phoretic model  \cite{Morozov2019JFM} it has been reported recently that for an infinite system size,
close to the bifurcation point from a nonmotile to a motile state, the velocity behaves as $v^0 \sim Pe-Pe_1$, and not as
$(Pe-Pe_1)^{1/2}$, as we found above. This result is confirmed by Saha et al. \cite{Saha2021}. Actually this  finding was also reported earlier by Rednikov
et al. \cite{rednikov1994drop} (see their equation 24). To be more precise $\lvert v^0 \rvert \sim  Pe-Pe_1$, and the absolute value is a signature of singular
behavior. The result $\lvert v^0 \rvert \sim Pe -Pe_1$ implies the existence of two symmetric branches of solutions $v^0 \sim \pm (Pe-Pe_1)$, and
the bifurcation is not transcritical, as stated in \cite{Morozov2019JFM}, but of pitchfork (albeit singular) nature. In simulations \cite{Hu2019} the size is finite and $v^0\sim (Pe-Pe_1)^{1/2} $ (pitchfork bifurcation). We have recently discussed\cite{Farutin_singular} how the singular bifurcation can be regularized (due, in particular, to a finite size) into a classical pitchfork bifurcation. 

Our study shows that the primary instability corresponds to a transition from a steady to a motile state with a straight trajectory.
That is, there is no possibility to go from a fully isotropic concentration distribution to a state with $v^0=0$ and $\omega^0\ne 0$.
This is because the rotational invariance of the concentration field needs to be broken first for solutions with time-dependent angular dependence to make sense.
Spiral-like trajectories have been observed in experiments.
A logarithmic spiral corresponds to a constant value of $\omega(t)$ and the amplitude of $v(t)$ that grows exponentially.
This suggests that spiral-like trajectories can not appear as a transient for a steady pitchfork bifurcation but can for Hopf bifurcation.
Detailed analysis of Hopf bifurcations in phoretic systems and finding a simple model which shows such behavior is thus a promising topic for future research.

Finally, the dynamics of a phoretic system does not necessarily relax to a self-congruent solution.
As the 2D studies show (see companion Letter \cite{misbahLetter}), periodic oscillations of strongly anharmonic form or even chaotic dynamics are among the possible solutions.
Investigating the 3D analogs of those types of dynamics represents another problem for future research.

\section{Acknowledgments}
We are grateful to Romy Morin for correcting several equations in the early version of the manuscript.
We thank CNES (Centre National d'Etudes Spatiales) (C.M. and A.F.) and the French-German university program "Living Fluids" (grant CFDA-Q1-14) (C.M., A.F. and S.R.). The computations presented in this paper were performed using the GRICAD infrastructure (\url{https://gricad.univ-grenoble-alpes.fr}), which is supported by Grenoble research communities.

\appendix
\section{\label{app1}Absence of swimming for a variational problem}
Here we show the absence of swimming for a variational problem. Suppose we have 
\begin{equation}
\label{evolution}
\dot c(x)=\mathcal G\{c(x')\}(x),
\end{equation}
where
\begin{equation}
\label{variational}
\mathcal G\{c(x')\}(x)=-\frac{\delta F\{c(x')\}}{\delta c(x)}
\end{equation}
for some non-linear functional $F$, and where $\delta F/\delta c$ refers to functional derivative.
Equation (\ref{variational}) can be rewritten, in view of the very definition of functional derivative, as
\begin{equation}
\label{variational2}
\lim_{h\rightarrow 0}\frac{F\{c(x)+h\delta c(x)\}-F\{c(x)\}}{h}=-\int \mathcal G\{c(x')\}(x)\delta c(x) dx.
\end{equation}

\subsection{Proof 1}
We can write that 
\begin{equation}
\label{invariance}
F\{c(x)\}=F\{c(x+h)\}=F\{c(x)+h\partial_{x}c(x)\}+O(h^2)
\end{equation}
because of the translational invariance of $F$.
Setting $\delta c$ to $\partial_x c$ in Eq. (\ref{variational2}) and comparing with eq. (\ref{invariance}) shows that
\begin{equation}
\label{zero}
\int \mathcal G\{c(x')\}(x)\partial_x c(x) dx=0.
\end{equation}
Multiplying Eq. (\ref{evolution}) by $\partial_x c$ and integrating with respect to x yields
\begin{equation}
\label{proof1}
\int \dot c(x)\partial_xc(x)dx=\int\mathcal G\{c(x')\}(x)\partial_xc(x)dx=0,
\end{equation}
Substituting the self-congruent solution $\dot c(x)=v\partial_x c(x)$ in Eq. (\ref{proof1}) shows that $v=0$ unless $\partial_x c(x)$ is zero everywhere.

\subsection{Proof 2}
The function $F\{c(x,t)\}$ is constant in time for a self-congruent solution due to the invariance of $F$ under rotations and translations of the concentration field $c$.
This expands to
\begin{equation}
\label{time}
\begin{aligned}
0=\frac{dF\{c(x,t)\}}{dt}&\equiv\lim_{h\rightarrow 0}\frac{ F\{c(x,t+h)\}-F\{c(x,t)\}}{h}\\
&=\lim_{h\rightarrow 0}\frac{F\{c(x,t)+h\dot c(x,t)\}-F\{c(x,t)\}}{h}=-\int \mathcal G\{c(x')\}(x) \dot c(x)dx=-\int \dot c(x)^2 dx,
\end{aligned}
\end{equation}
where the last equality uses (\ref{evolution}).
The right hand side of Eq. (\ref{time}) can be equal to zero only if $\dot c(x,t)=0$ for all $x$.
This shows that there are no motile self-congruent solutions of Eq. (\ref{evolution}) with variational right hand side.

\section{\label{Appendix}Extended analysis of the motility of a segment particle in 1D}
We write the time-dependent solution of eq. (\ref{functional}) as 
\begin{equation}
\label{solution}
c(x,t)=c_0(x-v^0t)+\varepsilon(t) f_1(x-v^0t)+\sum\limits_{i=2}^\infty \delta c_i(t) f_i(x-v^0t),
\end{equation}
where $\varepsilon(t)$ is a small parameter of the expansions and $\delta c_i(t)$ are small coefficients to be related to $\varepsilon$.
The velocity of the comoving frame $v^0$ is also small for $Pe$ close to $Pe_1$.

We express the amplitudes $\delta c_i$ as a function of $\varepsilon(t)$ by substituting (\ref{Gexpansion}) into the right hand side of eq. (\ref{functional}) and expanding the result in the basis $f_i$.

\begin{equation}
\label{dcidot}
\begin{aligned}
\dot {\delta c}_i&=G_1^{(i)}\{c(x')\}(Pe)+G_2^{(i)}\{c(x'),c(x'')\}(Pe)+o(\varepsilon,\delta c_j)\\
&=\lambda_i \delta c_i(t)+(Pe-Pe_1)\partial_{Pe} G_1^{(i)}\{c(x')\}(Pe_1)+G_2^{(i)}\{c(x'),c(x'')\}(Pe)+[o(\varepsilon,\delta c_j)]^2,
\end{aligned}
\end{equation}
where the $k$-linear functionals $G_k^{(i)}$ are defined by
\begin{equation}
\label{functionals}
\mathcal G_k(x,Pe)=\sum_{i=0}^\infty G_k^{(i)}(Pe)f_i(x,Pe_1)
\end{equation}
and we have made explicit the value of $Pe$ for which the functionals are evaluated.

The leading term of the right hand side of eq. (\ref{dcidot}) is $\lambda_i \delta c_i(t)$.
Since $\lambda_i$ is negative and has a large absolute value, the solution of (\ref{dcidot}) relaxes to a steady-state solution $\delta c_i^0$ on a time scale that is small compared to the dynamics of the $\varepsilon$ mode.
We find from this the steady-state value of $\delta c_i$ as
\begin{equation}
\label{dci0}
\begin{aligned}
\delta c_i^0&=-\frac{G_2^{(i)}\{c(x'),c(x'')\}(Pe_1)}{\lambda_2}+[o(\varepsilon,\delta c_j)]^2+O(Pe-Pe_1)O(\varepsilon,\delta c_j)\\
&=-\frac{G_2^{(i)}\{f_1(x'),f_1(x'')\}(Pe_1)}{\lambda_2}\varepsilon^2+o(\varepsilon^2)+O(Pe-Pe_1)O(\varepsilon),
\end{aligned}
\end{equation}
where the last equality is obtained by noting that $\delta c_j^0=O(\varepsilon^2)$, which is apparent from the first equality.
Note also that $\mathcal G_2\{f_1(x')f_1(x'')\}(x)$ is an even function of $x,$ so that $G_2^{(i)}\{f_1(x')f_1(x'')\}(x)=0$ if $f_i(x)$ is an odd function.
This means that $\delta c_i^0=O(\varepsilon^3)$ for such values of $i$.

Equation (\ref{dci0}) is obtained for a fixed value of $\varepsilon$ but it remains approximately valid if $\varepsilon$ is a slowly varying function of time.
In particular, it is valid if $\varepsilon$ saturates to a constant.
Taking the $f_1$ component of eq. (\ref{functional}) yields
\begin{equation}
\label{edot}
\begin{aligned}
\dot\varepsilon&=(Pe-Pe_1)\partial_{Pe}G_1^{(1)}\{c(x')\}(Pe_1)+G_2^{(1)}\{c(x'),c(x'')\}(Pe_1)+G_3^{(1)}\{c(x'),c(x''),c(x''')\}(Pe_1)\\
&=(Pe-Pe_1)\partial_{Pe}G_1^{(1)}\{f_1(x')\}(Pe_1)\varepsilon+\sum_{j=2}^\infty \left[G_2^{(1)}\{f_1(x'),f_j(x'')\}(Pe_1)+G_2^{(1)}\{f_j(x'),f_1(x'')\}(Pe_1)\right]\varepsilon\delta c_j^0\\
&+G_3^{(1)}\{f_1(x'),f_1(x''),f_1(x''')\}(Pe_1)\varepsilon^3+o(\varepsilon^3).
\end{aligned}
\end{equation}
Here we have used that $G_2^{(1)}\{f_1(x'),f_1(x'')\}(Pe_1)=0$, as discussed above.
Equation (\ref{edot}) is of form 
\begin{equation}
\label{pitchfork}
	\dot\varepsilon=a_1\varepsilon+a_3\varepsilon^3+o(\varepsilon^3),
\end{equation} which corresponds to a pitchfork bifurcation.
The transition point corresponds to $Pe=Pe_1$ and, assuming $\partial_{Pe}G_1^{(1)}\{c(x')\}(Pe_1)>0$, the stable steady-state value of $\varepsilon$ scales as $(Pe-Pe_1)^{1/2}$ for $Pe>Pe_1$.
Since $\lambda_1(Pe)<<|\lambda_j(Pe_1)|$ for $j>1$ and $|Pe-Pe_1|$ sufficiently small, the dependence of $\lambda_1$ on $Pe$ is governed by the $G_1^{(i)}\{f_j(x')\},$ $i,j\in \{0,1\}$ submatrix of $\mathcal G_1$.

Finally, expression for the translational velocity of the self-congruent solution (\ref{movesubs}) remains valid for $Pe\ne Pe_1$.
This completes the proof that $v^0\propto(Pe-Pe_1)^{1/2}$ regardless of the nature of the swimming particle as a general consequence of the translational and mirror symmetries of the system.
%
\end{document}